\documentclass[a4paper,12pt]{article}
\usepackage{srcltx}
\usepackage{times,amsbsy,amsmath,amsthm} 
\usepackage{graphicx,epstopdf}
\usepackage[hang]{subfigure}
\usepackage{multirow}
\usepackage{float}
\usepackage{hyperref}
\usepackage{cleveref}

\usepackage{color}
\usepackage{macros}

\newcommand\od[2]{\dfrac{\rd {#1}}{\rd {#2}}}
\newcommand\rd[0]{\mathrm{d}}
\newcommand{\pd}[2]{\dfrac{\partial #1}{\partial #2}}
\newcommand{\dd}{\,{\rm d}}

\newcommand\moment[1]{{\langle#1\rangle}}

\newtheorem{theorem}{Theorem}[section]
\newtheorem{lemma}{Lemma}[section]
\newtheorem{example}[theorem]{Example}
\newtheorem{definition}{Definition}

\title{
A Nonlinear Three-Moment Model for Radiative Transfer in Spherical Symmetry
}

\author{
Ruo Li\thanks{HEDPS \& CAPT, LMAM \& School of Mathematical
    Sciences, Peking University, Beijing, China, email: {\tt rli@math.pku.edu.cn}},~~
  Weiming Li\thanks{Applied and Computational Mathematics
    Division, Beijing Computational Science Research Center, Beijing, China
    and Department of Mathematics, The Hong Kong University of Science
\& Technology, Clear Water Bay, Kowloon, Hong Kong, China. email:
\tt{liweiming@csrc.ac.cn}}, 
  ~~and
Lingchao Zheng\thanks{School of Mathematical Sciences, Peking
    University, Beijing, China, email: {\tt lczheng@pku.edu.cn}} 
}

\begin{document}

\maketitle

\date{}


\begin{abstract}

We study the approximation of the radiative transfer equation with a
relatively few moments in the spherically symmetric case.  We propose
a three-moment model based on choosing the beta distribution as the
ansatz for the specific intensity. This ansatz enables our model to
capture the anisotropy in the distribution function. The
characteristic structure of the Riemann problem of the model is
studied in detail. Numerical simulations demonstrate its validity in
approximating the radiative transfer equation in the spherically
symmetric case and its advantage in approximating highly anisotropic
distribution functions in comparison to the $P_n$ method.

\end{abstract}




\section{Introduction}
\label{sec:B2_introduction}
The radiative transfer equation (RTE) is essential for describing the
propagation of radiation through absorbing and emitting
medium \cite{pomraning1973equations, mihalas2013foundations} and has
applications in the fields of astrophysics \cite{draine2010physics},
atmospheric physics \cite{marshak20053d} and optical imaging
\cite{klose2002optical}. It is a
high-dimensional integro-differential kinetic equation for the
specific intensity (also referred to as the distribution function in
this paper) of the radiation. Due to its high-dimensionality, numerical methods to
solve the RTE is a challenging topic.  Currently, numerical methods
for solving the radiative transfer equation can be categorized into
two groups: probabilistic approaches, for example, the direct simulation
Monte Carlo methods
\cite{lucy1999computing,hayakawa2007coupled,abdikamalov2012new}, and
the deterministic methods \cite{broadwell1964study, larsen2010advances,
dubroca1999theoretical, minerbo1978maximum,
alldredge2016approximating, fan2018fast}.  The 
deterministic methods mainly include the long or short characteristic methods
\cite{cannon1970line, kunasz1988short}, the discrete ordinate methods
($S_N$) \cite{jin1991discrete, sun2015asymptotic1}, and the moment
methods \cite{unno1989radiative, minerbo1978maximum}. 

The moment method is an effective approach to solve the radiative
transfer equation, as moments are often the quantities we are
interested in. For example, the radiation energy is the zeroth moment
and the radiation energy flux is the first moment. However, the
governing equation of each moment usually contains higher
order moments, therefore an approximation for the higher order
moments is required to complete the moment system.  This is called the
moment closure, which is the central problem of the moment
method.  Two important properties of the moment closure are
hyperbolicity and realizability. Hyperbolicity is a necessary
condition for the well-posedness of the first order moment systems,
while realizability is a natural requirement that comes from the
positivity of the underlying distribution function.  However, the most
commonly used moment model, the spherical harmonics ($P_N$) method
\cite{pomraning1973equations} does not possess realizability.
Therefore, the result of the $P_N$ model sometimes leads to negative
particle concentration \cite{brunner2001one,
brunner2005two,mcclarren2008solutions}.  In fact, for
multidimensional cases, the maximum entropy model ($M_N$)
\cite{dubroca1999theoretical, minerbo1978maximum} 
possesses both hyperbolicity and realizability. However,
the closing relationship of the maximum entropy model is not explicit,
making it too expensive to be applied in the numerical simulations. 

In some cases, the dimension of the radiative
transfer equation can be reduced due to symmetry considerations. One
widely studied dimension-reduced case is that of slab geometry
\cite{pomraning1973equations,
abdikamalov2012new, alldredge2016approximating, MPN}.
For the slab geometry case, the distribution function varies only along
one direction in spatial space and is an axisymmetric function in
microscopic velocity space. Another case is the spherically symmetric
case. The spherically symmetric case has many applications, 
such as in the studies of extended atmospheres
\cite{mihalas1978stellar}, 
and there are many previous studies  
\cite{gros1997implicit, anusha2009preconditioned,
kitzmann2016discontinuous}. Compared to the radiative transfer
equation in the slab geometry case, the transfer equation in spherical
symmetry is more complicated due to the appearance of a derivative
with respect to the microscopic velocity, 
which is not present in the slab geometry case.    

The present paper is an extension of a model previously studied in
\cite{alldredge2016approximating} for the slab
geometry case to the spherically symmetric case. The $B_2$ model
studied in \cite{alldredge2016approximating} is the second order case of the
Extended Quadrature Method of Moments (EQMOM)
\cite{abdikamalov2012new}. For the slab geometry cases,
the $B_2$ model gives an explicit closing relationship but possesses many of the essential properties of the
attractive but much more expensive maximum entropy model. 
It was shown that the $B_2$ model is
globally hyperbolic and possesses a non-negative ansatz for all realizable
moments, and numerical simulation shows it has similar results as the 
$M_2$ model for benchmark problems in slab geometry cases.

In this work, we apply the ansatz of the $B_2$ model to the
spherically symmetric case. Due to spherical symmetry, the
distribution function should also be axisymmetric, which indicates
that it is natural to extend the $B_2$ model to the spherically
symmetric case. The resulting
moment system is similar to the $B_2$ moment system in slab geometry
in the convection term and possesses both global hyperbolicity and realizability.
We also derive the connection condition between rarefaction and shock waves
when there are simple wave conditions. 
Several numerical simulations are computed to validate the $B_2$ 
model in spherically symmetric cases and show the advantages of the
$B_2$ model compared to other numerical methods. 

The rest of this paper is organized as follows: In Section
\ref{sec:preliminary} we briefly introduce the RTE in the spherically
symmetric case; 
then in \Cref{sec:B2_spherical_analysis} we introduce the $B_2$ model for
spherical symmetry and analyze its properties. Numerical results are
shown in \Cref{sec:numerics} and  
the paper ends with a brief summary and conclusion in \Cref{sec:conclude}.



\section{Preliminaries}
\label{sec:preliminary}
We consider the radiative transfer equation
(RTE) in spherical symmetry \cite{pomraning1973equations} 
\begin{equation}\label{eq:rt-spherical1D}
  \dfrac1c\pd{I}{t}+ \mu \pd{I}{r} + \dfrac{1-\mu^2}{r}\pd{I}{\mu} =
  \mathcal{C}(I),
\end{equation}
where
\begin{equation}
  \mathcal{C}(I) = 
 -\sigma_a I -\sigma_s I + 
  \dfrac{\sigma_s}{2} \int_{-1}^1 I(\mu)\dd\mu + \dfrac{S}{2}.
\end{equation}
Here $c$ is the speed of
light, $I=I(t,r,\mu)$ is the specific intensity, $t\in\bbR^+$ is time, 
$r\in[r_L,r_R]\subset \bbR^+$ is the
radial distance, and $\mu\in[-1,1]$ is the velocity related variable where
$\text{arccos}(\mu)$ represents the polar angle, i.e., the angle
between the photon velocity and the radial distance $\br$. Also, $\siga$ and
$\sigs$ are the absorption and scattering coefficients, respectively,
and $S$ is the external source term. Compared to the radiative
transfer equation in slab geometry, which has the form
\begin{equation}\label{eq:rt-slab}
  \dfrac1c \pd{I}{t} + \mu \pd{I}{x} = \mathcal{C}(I),
\end{equation}
equation \eqref{eq:rt-spherical1D} is more
complicated because of the presence of a derivative with respect to the polar
angle, a term which does not appear in \eqref{eq:rt-slab}.

We can obtain a moment system from equation \eqref{eq:rt-spherical1D}
by taking its moments with respect to $\mu$, 
truncating the system and specifying a closure. Denote
the $k$-th moment of the specific intensity as
\begin{equation}
\langle I \rangle _k \triangleq \int_{-1}^1 \mu^k I(\mu)\dd\mu,\quad k \in \bbN,
\label{eq:momentsdefine}
\end{equation}
we multiply equation \eqref{eq:rt-spherical1D} by $\mu^k$ and
integrate over $\mu \in [-1, 1]$ to get  
\begin{equation}
  \label{eq:momentspherical}
  \begin{aligned}
    \dfrac{1}{c} \pd{\vint{I}_k}{t} 
    +\pd{\vint{I}_{k+1}}{r} +
    \dfrac{(k+2) \vint{I}_{k+1}-k \vint{I}_{k-1}}{r} 
  = \vint{\mathcal{C}(I)}_k, 
\end{aligned}
\end{equation}
where
\begin{equation}
\vint{\mathcal{C}(I)}_k = -(\sigma_a + \sigma_s)
  \vint{I}_{k} + \dfrac{1+(-1)^k}{2k+2}(\sigma_s \vint{I}_{0}+S).
\end{equation}
The governing equation of $\vint{I}_k$ in \eqref{eq:momentspherical}
depends on higher order moment $\vint{I}_{k+1}$, indicating that the
full system contains an infinite number of equations. For a given integer
$N \in \bbN$, we truncate the system and only keep the governing
equations of $\vint{I}_k$, $k \leq N$. However, the first $N+1$
equations contain $N+2$ unknown variables; thus the truncated system
is not closed. In order to obtain a closed system, we have to
approximate the $(N+1)$-th moment $\vint{I}_{N+1}$ based on the first
$N+1$ moments $\vint{I}_k$, $k = 0, \cdots, N$, which is formulated as 
\[ 
  \vint{I}_{N+1} \approx E_{N+1} = E_{N+1}(\vint{I}_0,
  \vint{I}_1,\cdots, \vint{I}_N),
\]
and this is called the moment closure. A popular way to obtain the
moment closure is to construct a specific intensity $\hat{I}$
based on the known
moments. Let $E_k$, $k=0,1,\cdots,N$, be the known moments of a
specific intensity $I$, we can construct a specific intensity
$\hat{I}(\mu;E_0,E_1,\cdots,E_N)$, which is called the ansatz of $I$,
to approximate the unknown specific intensity $I$, such that,
\begin{equation}\label{eq:momentsconstrations}
  \int_{-1}^1 \mu^k\hat{I}(\mu;E_0,E_1,\cdots,E_N) \dd\mu = E_k, \quad
  k=0,1,\cdots,N.
\end{equation}
Then the $(N+1)$-th moment can be directly approximated by the $(N+1)$-th
moment of $\hat{I}$, 
\[ 
  E_{N+1} = \int_{-1}^1
  \mu^{N+1}\hat{I}(\mu;E_0,E_1,\cdots,E_N)\dd\mu,
\]
which closes the moment system \eqref{eq:momentspherical}.  

The physically relevant value of a given set of moments for the RTE
should satisfy constraints imposed by the positivity of the specific
intensity. It is referred to as the realizable region. Realizability
of the closing relationship ensures the closure is consistent with our
apriori knowledge about the moments of a probability distribution
function and is expected to improve the accuracy of the truncated
system of equations \cite{kershaw1976flux}.  Below we recall the
definition of the realizable region.
\begin{definition} The realizable region for the first $N+1$ moments
  for any given integer $N \in \bbN$, denoted as $\Omega_{R,N}$, is
  the set of moments where each point corresponds to a positive
  specific intensity,  
  \[ 
    \Omega_{R,N} \triangleq \{ (E_0,E_1, \cdots, E_N)^T : \exists I(\mu)> 0,
    \moment{I}_k = E_k, k=0,1,\cdots,N \}.
  \]
\end{definition}

The focus of this paper is to find a closing relationship of the second
order moment model, for which the known moments are in the realizable region,
which we will study in the next section.



\section{$B_2$ Model for Spherical Symmetry}
\label{sec:B2_spherical_analysis}
\subsection{The $B_2$ closure}
The moment model for $N = 2$ is written as 
\begin{equation}\label{eq:moment-model-spherical-2nd}
  \dfrac1c \pd{\bE}{t} + \pd{\bF}{r} = \bC,
\end{equation}
where $\bE=(E_0,E_1,E_2)^T$, $\bF=(E_1,E_2,E_3)^T$, and 
\[
\bC_k = -\frac{(k+2)E_{k+1}-kE_{k-1}}{r}-\sigma_t E_k
+\frac{1+(-1)^k}{2k+2}(\sigma_s E_0+ S), \quad k=0,1,2.
\]

The realizable region for $(E_0, E_1, E_2)$ can be derived from the
Cauchy-Schwarz inequality to be
\begin{equation} \label{eq:realizablityspace}
  \Omega_{R,2} = \{(E_0,E_1,E_2)^T : E_0>0, E_2< E_0, E_1^2 < E_0 E_2\}.
\end{equation}

To close system \eqref{eq:moment-model-spherical-2nd}, we use the
following beta distribution ansatz,
\begin{equation}\label{eq:ansatz-B2}
  \bansatz(\mu;E_0,E_1,E_2) = 
	\dfrac{w}{2 \bB(\xi, \eta)} \left(\dfrac{1 +
  \mu}{2}\right)^{\xi - 1} \left( \dfrac{1-\mu}{2}\right)^{\eta - 1},
  \xi = \dfrac{\gamma}{\delta}, ~~\eta = \dfrac{1-\gamma}{\delta},
\end{equation}
with $w>0, 0<\gamma<1,\delta>0$. The ansatz \eqref{eq:ansatz-B2} has the same form as the $B_2$ ansatz
for slab geometry previously studied in
\cite{alldredge2016approximating}.  
Consistency of the known moments \eqref{eq:momentsconstrations}
determines the parameters $w$, $\gamma$, and $\delta$, as
\begin{equation}
 w = E_0, \quad
 \gamma = \frac{E_1 / E_0 + 1}2, \quand 
 \delta = \frac{(E_1 / E_0)^2 - E_2 / E_0}{E_2 / E_0 - 1},
\end{equation}
giving the moment closure
\begin{equation}\label{eq:beta-closure}
  E_3 = \int_{-1}^1 \mu^3 \bansatz  \dd \mu
  =\frac{E_1(E_2^2 + 2E_1^2 - 3E_0 E_2)}{2E_1^2 - E_0 E_2 - E_0^2}.
\end{equation}

System
\eqref{eq:moment-model-spherical-2nd} with the closing relationship
\eqref{eq:beta-closure} shall be referred to as the $B_2$ model for
spherical symmetry hereafter, and it differs from that studied in
\cite{alldredge2016approximating} in that it contains curvature terms
brought about by spherical geometry.  

In \cite{alldredge2016approximating} it was proved that the $B_2$ model
for slab geometry is globally hyperbolic, and there exists a
non-negative ansatz for realizable moments. Hyperbolicity is necessary
for the well-posedness of the Cauchy initial value problem. 
Moreover, 
the existence of a non-negative ansatz ensures the realizability of the
closure. 
System \eqref{eq:moment-model-spherical-2nd} has the
same convection terms as the $B_2$ system for slab geometry studied in
\cite{alldredge2016approximating}, therefore it is also globally
hyperbolic and realizable. 

Next, we study the Riemann problem because its solution structure is
instructive for the design of the approximate Riemann solver, which
is the basis of the numerical methods using Godunov-type schemes. Note
that this
problem has not yet been studied for the slab geometry case, and the
analysis below can also be applied to that scenario.

\subsection{Riemann problem}
Define $\bE = (E_0, E_1, E_2)^T$ and $\bF = (E_1,E_2,E_3)^T$, where $E_3$
is given by \eqref{eq:beta-closure}.  Consider the following Riemann
problem:
\begin{equation}
  \begin{split}
    & \dfrac{1}{c}\pd{\bE}{t} + \pd{\bF}{r} = 0, \\
    & \bE(0, r) = \left\{\begin{array}{l}
        \bE^L, r < r_0, \\
        \bE^R, r > r_0. \\
      \end{array}\right.
  \end{split}
\end{equation}

We first recall results from \cite{alldredge2016approximating} in
two lemmas which will be used in the following discussions in this
section. Hereafter, the eigenvalues and eigenvectors of the Jacobian matrix
\begin{equation}\label{eq:Jacobi-of-ApproxM2}
  \boldsymbol{J} = \pd{\bF}{\bE} =\left(\begin{array}{ccc}
      0 & 1 & 0\\
      0 & 0 & 1\\
      \pd{E_3}{E_0} & \pd{E_3}{E_1}& \pd{E_3}{E_2}\\
    \end{array}
  \right),
\end{equation}
are denoted by $\lambda_k$ and $\br_k$, $k=1,2,3$, with
$\lambda_1<\lambda_2<\lambda_3$, and we also define $\Delta_k
\triangleq \nabla_{\bE} \lambda_k \cdot \br_{k}$, which is essential
to analyze the characteristic structure. 
\begin{lemma}
  \label{lem:lambdaandE}
  For any $\bE\in\Omega_{R,2}$, we have 
  \( 
     -1<\lambda_1<\dfrac{E_1}{E_0}<\lambda_3<1.
  \)
\end{lemma}
Moreover, the eigenvector of $\bJ$ with respect to the eigenvalue $\lambda_k$ is
$ \br_{k}=(1,\lambda_k,\lambda_k^2)^T.$
\begin{lemma}
	\label{lem:delta_lambda} 
    $\Delta_1<0$ and $\Delta_3>0$ hold for any $\bE\in\Omega_{R,2}$. However, 
    $\Delta_2$ changes sign over $\Omega_{R,2}$.
\end{lemma}

It was proved in \cite{alldredge2016approximating} that the first and third
characteristic fields are genuinely nonlinear, while the second
characteristic field is neither genuinely nonlinear nor linearly
degenerate. This indicates the solution of the Riemann problems may
contain rather complex behavior. However, we would still expect
simple characteristic waves in the solution of Riemann problems for
first and third waves.  Simple wave solutions of the above Riemann
problems are the composition of one intermediate state connected by
rarefaction waves or shock waves. In order to get full understanding
of the system, these waves will be studied respectively below.  

To simplify discussions, we denote 
\[
  u=\dfrac{E_1}{E_0},\quad 
  p=E_2-\dfrac{E_1^2}{E_0}.
\]

\paragraph{Rarefaction waves}
We first investigate the integral curves given by rarefaction waves.
We take the first characteristic field as an example.
Suppose that two states $\bE^L=(E_0^L,E_1^L,E_2^L)^T$ and
$\bE^R=(E_0^R,E_1^R,E_2^R)^T$ are connected by a rarefaction wave
corresponding to the first characteristic field. Take $\zeta$ as the
parameter such that $\tilde{\bE}(0)=\bE^L$ and
$\tilde{\bE}(\zeta_*)=\bE^R$, while 
\begin{equation}
  \tilde{\bE}(\zeta) = \left( \tilde{E}_0(\zeta), 
  \tilde{E}_1(\zeta), \tilde{E}_2(\zeta)
  \right)^T
\end{equation}
denotes the integral curve in the $3$-dimensional phase space. The
corresponding eigenvalues and eigenvectors are denoted as
$\tilde{\lambda}_1(\zeta)$ and $\tilde{\br}_1(\zeta)$.

According to the theory of integral curves of rarefaction waves 
\cite[Chapter 13.8.1]{leveque2002finite}, there exists a 
nonzero scalar factor $\tilde{\alpha}(\zeta)$, such that 
\begin{equation}\label{eq:integral-curve}
  \tilde{\bE}'(\zeta) = \tilde{\alpha}(\zeta)
  \tilde{\br}_1 (\tilde{\bE}(\zeta)).
\end{equation}
Recalling $\tilde{\br}_1 = (1, \tilde{\lambda}_1,
\tilde{\lambda}_1^2)^T$, 
by direct calculation we obtain from \eqref{eq:integral-curve} 
\begin{equation}
  \label{eq:characteristicdeduction}
  \od{\tilde{E}_0(\zeta)}{\zeta} = \tilde{\alpha}(\zeta),\quad 
    \od{\tilde{u}(\zeta)}{\zeta} =
    \tilde{\alpha}(\zeta)(\tilde{\lambda}_1-\tilde{u}),\quad 
    \od{\tilde{p}(\zeta)}{\zeta}=\tilde{\alpha}(\zeta)(\tilde{\lambda}_1-\tilde{u})^2.
  \end{equation}

  Combine the entropy condition
$\tilde{\lambda}_1(0)=\lambda_1^L < \lambda_1^R=
\tilde{\lambda}_1(\zeta_*)$ 
with
\[
\dfrac{1}{\tilde{\alpha}
(\zeta)}\od{\tilde{\lambda_1}(\zeta)}{\zeta} = 
\dfrac{1}{\tilde{\alpha}(\zeta)}
\nabla_{\tilde{\bE}}\tilde{\lambda}_1\cdot \od{\tilde{\bE}}{\zeta}=
\nabla_{\tilde{\bE}}\tilde{\lambda}_1\cdot\tilde{\br}_1
	<0,
\]
where the last inequality holds because of \Cref{lem:delta_lambda},
we could see that $\tilde{\alpha}(\zeta_*)\zeta_* < 0$. 
Also, according to \Cref{lem:lambdaandE} we have
  $\tilde{\lambda}_1<\tilde{u}$, therefore
  \eqref{eq:characteristicdeduction} indicates that  
  \[
      E_0^L>E_0^R,\quad u^L<u^R,\quad p^L>p^R.
  \]

Analogously, for the third wave, we have
\begin{equation}
  E_0^L < E_0^R, \quad u^L < u^R, \quad p^L < p^R.
\end{equation}

\paragraph{Shock waves}
We consider shock waves by taking the first characteristic
field as an example. Suppose $\bE^L=(E_0^L,E_1^L,E_2^L)^T$ and
$\bE^R=(E_0^R,E_1^R,E_2^R)^T$ are connected by a shock corresponding
to the first characteristic field, then according to the
Rankine-Hugoniot condition, there exists
$s_1\in\bbR\backslash \{0\}$, such that 
\begin{equation} \label{eq_shock_RH}
    \begin{aligned}
        E_1^R-E_1^L &= s_1 (E_0^R-E_0^L),\\
        E_2^R-E_2^L &= s_1 (E_1^R-E_1^L),\\
        E_3^R-E_3^L &= s_1 (E_2^R-E_2^L).
    \end{aligned}
\end{equation}
From the first two equations of \eqref{eq_shock_RH},    
\begin{equation}\label{eq:shock_sk}
    \begin{aligned}
    s_1 &= \frac{E_1^R-E_1^L}{E_0^R-E_0^L} = \frac{E_2^R-E_2^L}{E_1^R-E_1^L}.
    \end{aligned}
\end{equation}
Direct calculations from the above equalities yield 
\begin{equation}\label{eq:shock_up}
  \begin{aligned}
  &(E_0^R-E_0^L)(u^R-u^L) =
  \dfrac{(E_0^R-E_0^L)^2}{E_0^R}(s_1-u^L),\\
    &(E_0^R-E_0^L) (p^R-p^L) = 
    \dfrac{1}{E_0^LE_0^R}(E_0^RE_1^L-E_0^LE_1^R)^2 > 0.
\end{aligned}
\end{equation}
Also, according to the entropy condition and \Cref{lem:lambdaandE},  
\begin{equation}\label{eq:entropycondition}
    u^L>\lambda_k^L>s_1>\lambda_k^R,
\end{equation}
indicating $(E_0^R-E_0^L)(u^R-u^L)< 0$. Therefore, the signs of $u^R-u^L$ 
and $p^R-p^L$ are determined by sign of $E_0^R-E_0^L$.

Next, we consider the sign of $E_0^R-E_0^L$. Parameterize the Hugoniot curve as
$\tilde{\bE}(\tau)$ with $\tilde{\bE}(0)=\bE^L$ 
and $\tilde{\bE}(1)=\bE^R$. 
For any $0\leq
\tau_*<\tau_*+\varepsilon\leq 1$ define $\bE^l=\tilde{\bE}(\tau_*)$ and 
$\bE^r=\tilde{\bE}(\tau_*+\varepsilon)$, then $\bE^l$ and $\bE^r$ satisfy the Rankine-Hugoniot condition
\eqref{eq_shock_RH} and the entropy condition
\eqref{eq:entropycondition}, therefore also \eqref{eq:shock_up}.
Assuming $\varepsilon$ is sufficiently small, then according to
\eqref{eq_shock_RH}, we have 
\[
   E_0^r=E_0^l + d,\quad E_1^r=E_1^l + s_1 d,\quad 
	 E_2^r=E_2^l + s_1^2d, \quad E_3^r=E_3^l + s_1^3 d,
\]
where $|d|\ll 1$, and $\lambda_1^l>s_1>\lambda_1^r$. As
$\lambda_1$ is a smooth function of $\bE$, Taylor expansion yields 
\begin{align}
  \label{eq:characteristicdeduction2}
\lambda_1^r-\lambda^l_1 = \lambda_1(\bE^r)-\lambda_1(\bE^l)=d\left(
\left(\pd{\lambda_1}{E_0}\right)^l+\left(\pd{\lambda_1}
{E_1}\right)^ls_1
+\left(\pd{\lambda_1}{E_2}\right)^ls_1^2
\right)  + O(d^2).
\end{align}
According to Lemma \ref{lem:delta_lambda}, we have  
\begin{align} 
\left(\pd{\lambda_1}{E_0}\right)^l+\left(\pd{\lambda_1}
{E_1}\right)^l\lambda_1^l
+\left(\pd{\lambda_1}{E_2}\right)^l(\lambda^l_1)^2 =\Delta_1 < 0.
\end{align}
Notice that  $|d|\ll 1$ indicates
$|\lambda_1^r-\lambda_1^l|\ll 1$, thus $|s_1-\lambda_1^l|\ll 1$.
Therefore, when $\vert d\vert \ll 1$, 
\begin{align} 
  \label{eq:characteristicdeduction3}
\left(\pd{\lambda_1}{E_0}\right)^l+\left(\pd{\lambda_3}
{E_1}\right)^l s_1
+\left(\pd{\lambda_1}{E_2}\right)^l(s_1)^2 < 0. 
\end{align}
Combining \eqref{eq:characteristicdeduction2} and 
\eqref{eq:characteristicdeduction3} with 
entropy condition \eqref{eq:entropycondition}, we obtain 
\begin{equation}
  \label{eq:analysis}
              d > 0,\quad  E_0^r > E_0^l,
							\quad u^r < u^l, \quad p^r > p^r.
\end{equation}

Notice that \eqref{eq:analysis} holds for any $\tau_*\in[0,1)$, and
according to the finite covering theorem, we
can obtain the following results for the first characteristic field. 
\begin{equation}
    E_0^R > E_0^L,\quad 
    u^R < u^L,\quad 
    p^R > p^L.
\end{equation}

For the third characteristic field, we analogously have
\begin{equation}
    E_0^R < E_0^L,\quad 
    u^R < u^L,\quad 
    p^R < p^L.
\end{equation}

Before the end of this subsection, we summarize 
the discussion on the entropy condition of two types of waves
in the following theorem.
\begin{theorem}
    For the $B_2$ moment system, the waves of the first and third
    families are elementary, and their types can be determined by the
    sign of $\lambda - u$ and the macroscopic parameters on the
    both sides of the waves:
\begin{table}[htbp]
    \centering
    \begin{tabular}{|r|c|c|c|}
      \hline 
      & $E_0$   & $u$ & $p$ \\
      \hline 
      \multirow{2}{*}{Rarefaction wave}  &   
      1-wave $E_0^L>E_0^R$  &
      \multirow{2}{*}{$u^L < u^R$} & 
      1-wave $p^L>p^R$ 
      \\  
      \cline{2-2}
      \cline{4-4}
      &  3-wave $E_0^L<E_0^R$
      & & 3-wave $p^L<p^R$ 
      \\
      \hline   
      \multirow{2}{*}{Shock wave}  &   
      1-wave $E_0^L<E_0^R$  &
      \multirow{2}{*}{$u^L > u^R$} & 
      1-wave $p^L<p^R$ 
      \\  
      \cline{2-2}
      \cline{4-4}
      & 3-wave $E_0^L>E_0^R$
      & & 3-wave $p^L>p^R$ 
      \\
      \hline 
    \end{tabular}
    \caption{wave structure of one dimension $B_2$ model}
  \end{table}
\end{theorem}


\section{Numerical Experiments}
\label{sec:numerics}
In this section, we discuss the numerical scheme for the $B_2$ model
for spherical symmetry and perform numerical
simulations on some typical examples to verify its validity. In all
subsequent computation we assume $c = 1$ in equation
\eqref{eq:moment-model-spherical-2nd}.

\subsection{Numerical scheme}

We use an equidistant discretization of $N_{\text{cell}}$ cells 
 for the spatial domain $[r_L,r_R]$, where $0\leq r_L<r_R$. Therefore,
 the spatial step is $\Delta r =
 \frac{r_R-r_L}{N_{\text{cell}}}$, with grid points $r_i=r_L+(i-1/2)\Delta
 r,i=1,\cdots,N_{\text{cell}}$, and cells $[r_{i-1/2},r_{1+1/2}]$,
 $i=1,\cdots,N_{\text{cell}}$, with midpoints
 $r_{1-1/2}=r_i-\Delta r/2$. We denote the approximation of the
 solution and the source term 
 in the $i$-th cell at time step $t_n$ as $\bE_i^n$ and $\bC_i^n$,
 respectively. 

 We adopt the splitting scheme 
described in \cite[Chapter
17.6]{leveque2002finite}. Problem
\eqref{eq:moment-model-spherical-2nd} is split into a
hyperbolic conservation law, $\pd{\bE}{t} + \pd{\bF}{r} = 0$, and a
simple ODE, $\pd{\bE}{t} = \bC$. The former is solved under the finite volume
framework, and the latter is solved
using the backward Euler method. The numerical scheme is formulated as 
\begin{align}
  \label{eq:numericalschemedetail}
  \begin{aligned}
  &\dfrac{\bE_{i}^{*}-\bE_i^n}{\Delta t} + 
  \frac{\bF_{i+1/2}-\bF_{i-1/2}}{\Delta r_i} = 0,\\
  &\dfrac{\bE_{i}^{n+1}-\bE_i^*}{\Delta t} = \bC_{i}^{n+1},
  \end{aligned}
\end{align}
where $\bE_{i}^*$ is the intermediate value in the $i$-th cell between
the time step $t_n$ and $t_{n+1}$, 
$\bF_{i+1/2}$ is the numerical flux at $r=r_{i+1/2}$,
and the $k$-th entry of the
source term $\bC_i^{n+1}$ is  
\[ 
C_{k,i}^{n+1} = -\dfrac{(k+2)E_{k+1,i}^{n+1}-kE_{k-1,i}^{n+1}}{r_i}
-\sigma_{t,i}^{n+1}E_{k,i}^{n+1}
+\dfrac{1+(-1)^k}{2k+2}(\sigma_{s,i}^{n+1}E_{0,i}^{n+1} +
S_{i}^{n+1}).
\]
We compute $\bF_{i+1/2}$ using the Lax-Friedrich flux and solve the
nonlinear system resulting from the implicit discretization of the source term by
the Newton iteration method. 

The time step is constrained by
the CFL condition 
\[
   \Delta t = \text{CFL} \cdot \min_i \dfrac{\Delta r}{\max_k
   |\lambda_k(\bE_i^n)|},
\] 
where $\lambda_k(\bE_i^n)$ is the $k$-th characteristic value of the
$B_2$ model when the given moments are $\bE_i^n$.  Notice that it is
proved in \cite{alldredge2016approximating} that the characteristic
speed of the $B_2$ model is not larger than 1.  Hence we take $\Delta
t = \text{CFL} \cdot \Delta r$, and the CFL number is chosen to be 0.95.

In our numerical scheme, we also need to specify the values of
$E_1$, $E_2$, and $E_3$ on the boundaries, which we 
derive from the kinetic level. Take the left boundary as an example.  
We define the flux vector $\hat \bff^{\rm L} := (E^{\rm
L}_1, E^{\rm L}_1, E^{\rm L}_3)^T$ on the left boundary as
$$
E^{\rm L}_j = \int_0^1 \mu^j I_{\text{out}} d\mu + 
    \int_{-1}^0 \mu^j \bansatz d\mu, \quad \text{for} ~~j=1,2,3,
$$
where $\bansatz$ is the ansatz associated with the moment vector from
the first spatial cell on the inner side of the boundary and  
$I_{\text{out}}$ is the distribution function outside of the domain
depending on $\bansatz$ and the actual boundary condition of the problem.
For instance, for the vacuum boundary condition, 
\[ 
   I_{\text{out}}(t,r_L,\mu) = 0,\quad \mu>0,
\]
while for the reflective boundary condition, 
\[
    I_{\text{out}}(t,r_L,\mu) =  \bansatz(t,r_L,-\mu),\quad \mu>0.
\]
We use the reflective boundary condition on the left when $r_L=0$.
Analogously, we can give the boundary condition for the right boundary
at $r_R>0$.

\subsection{Numerical results}
\label{sec:numerical_results}
\begin{example}[Bilateral beams]
  \label{exm:example1}
This example is used to test the ability of the $B_2$ model to deal
with anisotropic specific intensity.  
In this test, the spatial domain is unbounded (but $[0,1]$ for
numerical experiments), $\siga$ and  $\sigs$
are set to 0, and the external source term $S=0$. 
The initial state is taken as  
\[
   I(0,r,\mu) = \left\{
   \begin{aligned}
     &\dfrac{1}{r^2}\delta(\mu-1), \quad &0.2\leq r\leq 0.4,\\
     &\dfrac{1}{r^2}\delta(\mu+1), \quad &0.7\leq r\leq 0.9,\\
     &0,\quad & \text{otherwise}.
   \end{aligned}
   \right.
\]
\end{example}
The analytical solution of this problem is  
\[ 
  I(t,r,\mu) = 
  \left\{
  \begin{aligned}
    &\dfrac{(r-t)^2}{r^2}I(0,r-t,1),\quad & \mu=1,\\
    &\dfrac{(r+t)^2}{r^2}I(0,r+t,-1),\quad & \mu=-1.
\end{aligned}
\right.
\]
Therefore, the analytical solutions of $E_0$, $E_1$ and $E_2$ are 
\[
E_0(t,r)= E_2(t,r)= \dfrac{1}{r^2} \chi_{[0.2+t,0.4+t]}(r)
+\dfrac{1}{r^2} \chi_{[0.7-t,0.9-t]}(r),
\]
\[
E_1(t,r) = \dfrac{1}{r^2} \chi_{[0.2+t,0.4+t]}(r)
-\dfrac{1}{r^2} \chi_{[0.7-t,0.9-t]}(r),
\]
where $\chi_A(r)$ is the indicator function, defined as  
\[ 
 \chi_A(r) = \left\{
 \begin{aligned}
   &1,\quad &r\in A,\\
   &0,\quad &r\notin A.
 \end{aligned}
 \right.
\]
This problem describes two shells. The inner shell spreads out while
the outer shell contracts to the center. The specific intensities of
these two shells are delta functions.  When the two shells intersect,
the specific intensity is the sum of two delta functions
$\delta(\mu-1)$ and $\delta(\mu+1)$.  This kind of highly anisotropic
specific intensity  is difficult to approximate by traditional methods
\cite{MPN}. 

However, 
when the actual distribution function is the sum of
two delta functions, 
\begin{equation}
  \label{eq:distributionexample}
   I = C_1\delta(\mu-1) + C_2\delta(\mu+1)
\end{equation}
where $C_1$ and $C_2$ are constants.
The moments of the distribution function in
\eqref{eq:distributionexample} are $E_0=E_2=C_1+C_2$ and $E_1=C_1-C_2$.
According to the closure
\eqref{eq:beta-closure}, we have that the moment closure given by the $B_2$ model 
is 
\[  
  E_3 = C_1-C_2,
\]
which is the third moment of the actual distribution function.
Therefore, the $B_2$ model can deal with this kind of highly
anisotropic distribution function.

\begin{figure}[htbp]
    \centering
    \subfigure[solution of $E_0$ at $t = 0$.]{  
    \includegraphics[width=0.3\textwidth,height=0.16\textheight]{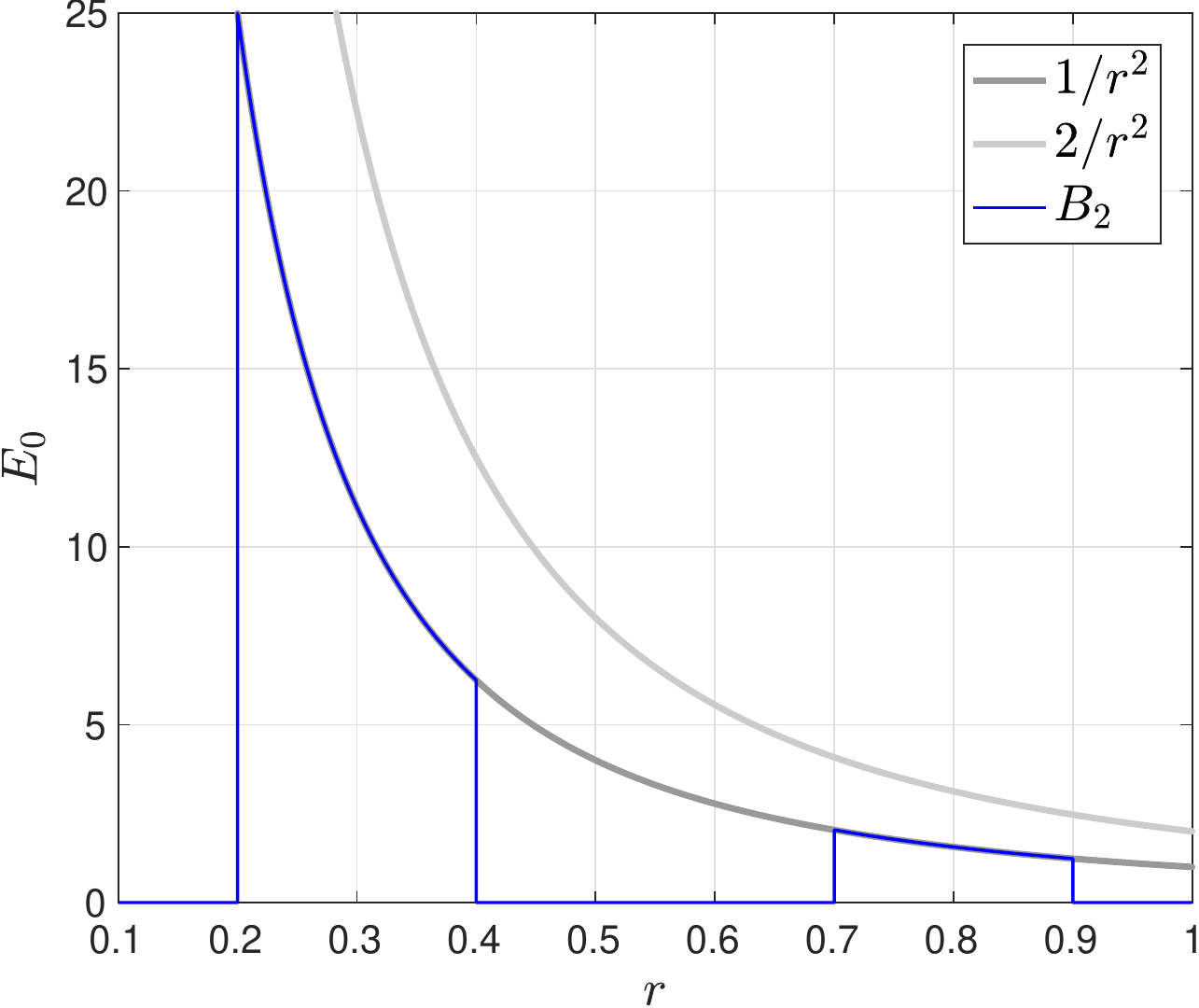}}
    \subfigure[solution of $E_0$ at $t = 0.1$.]{  
    \includegraphics[width=0.3\textwidth,height=0.16\textheight]{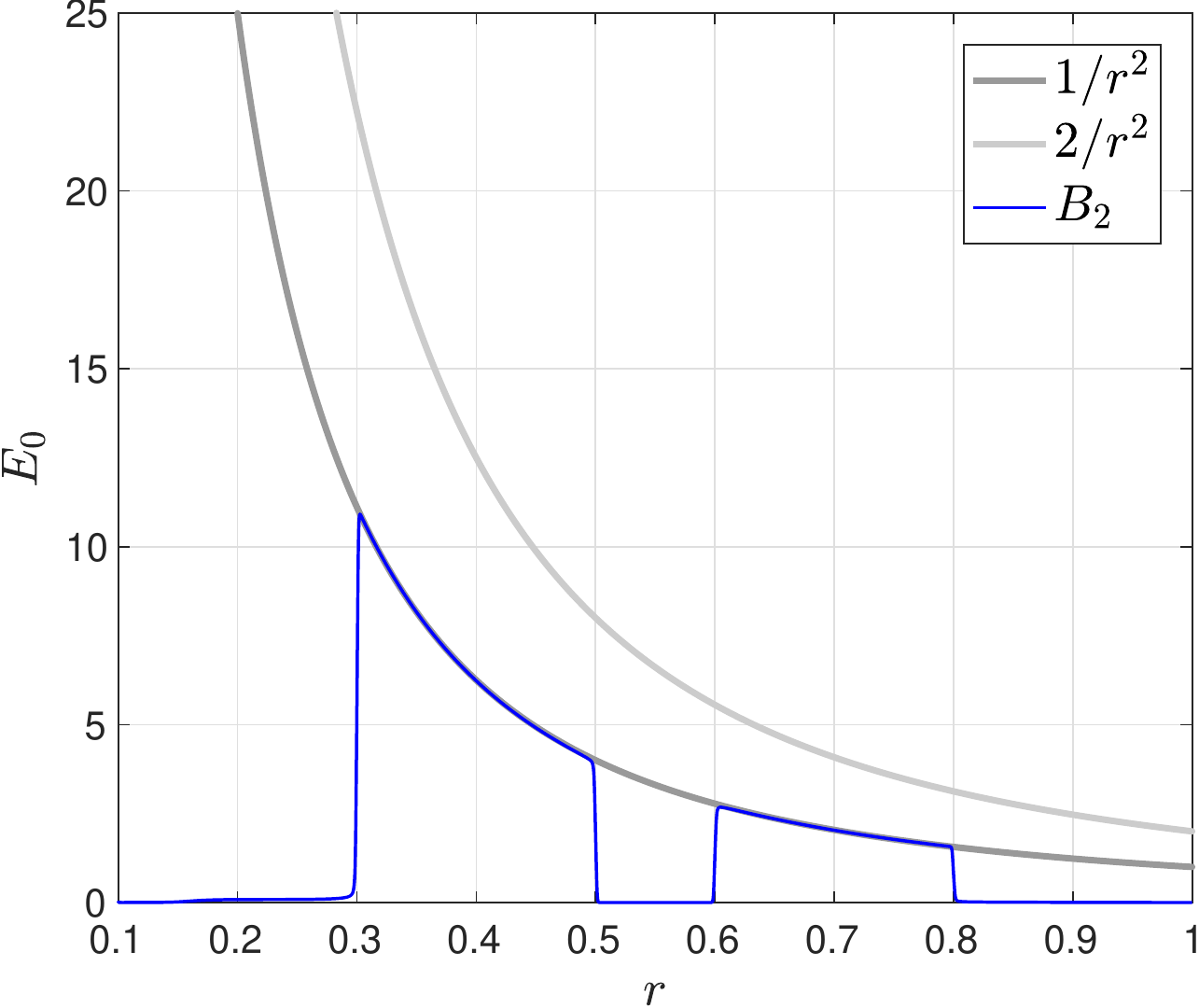}}
    \subfigure[solution of $E_0$ at $t = 0.2$.]{  
    \includegraphics[width=0.3\textwidth,height=0.16\textheight]{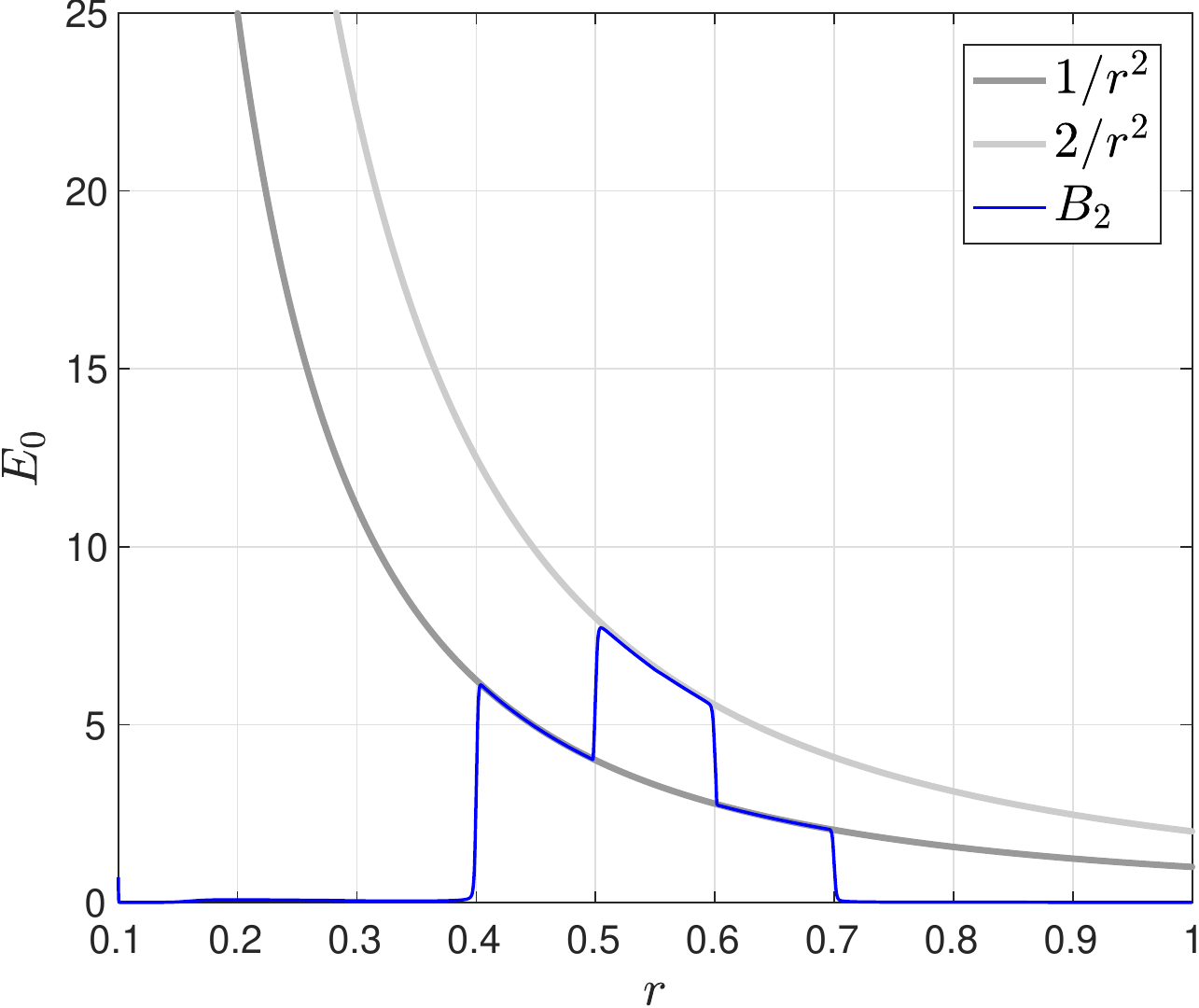}}
    \hfill
    \subfigure[solution of $E_0$ at $t = 0.25$.]{  
    \includegraphics[width=0.3\textwidth,height=0.16\textheight]{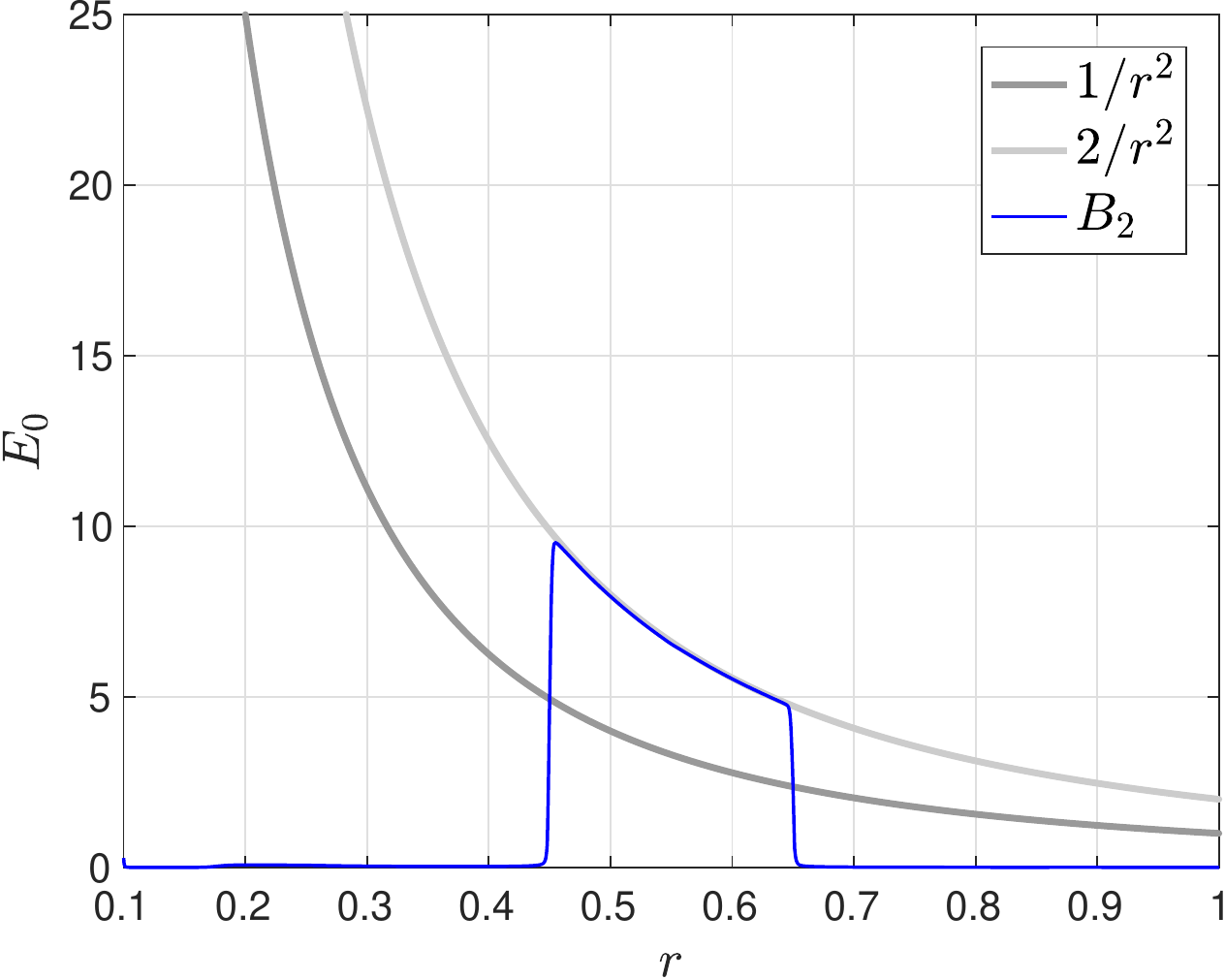}}
    \subfigure[solution of $E_0$ at $t = 0.3$.]{  
    \includegraphics[width=0.3\textwidth,height=0.16\textheight]{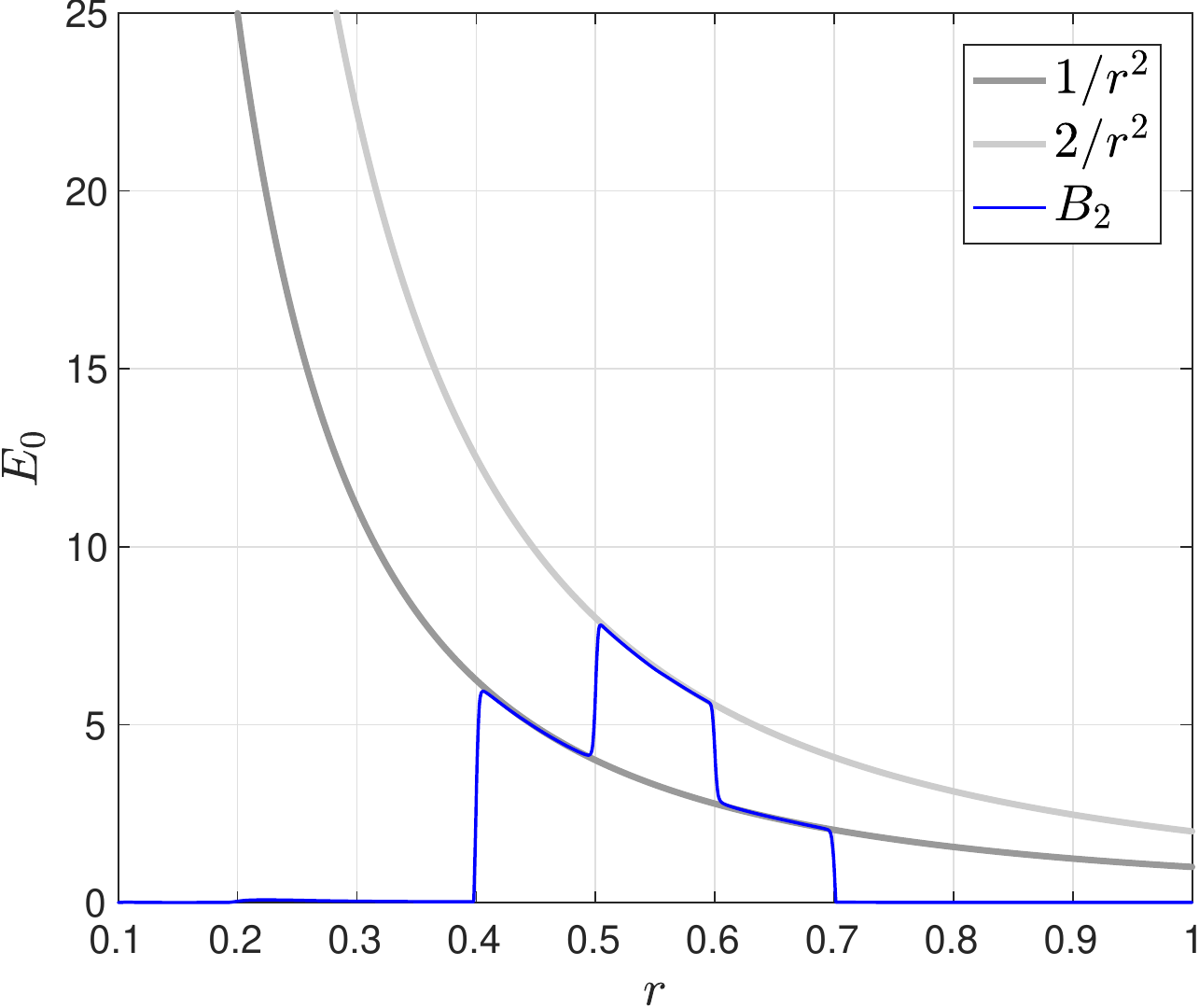}}
    \subfigure[solution of $E_0$ at $t = 0.4$.]{  
    \includegraphics[width=0.3\textwidth,height=0.16\textheight]{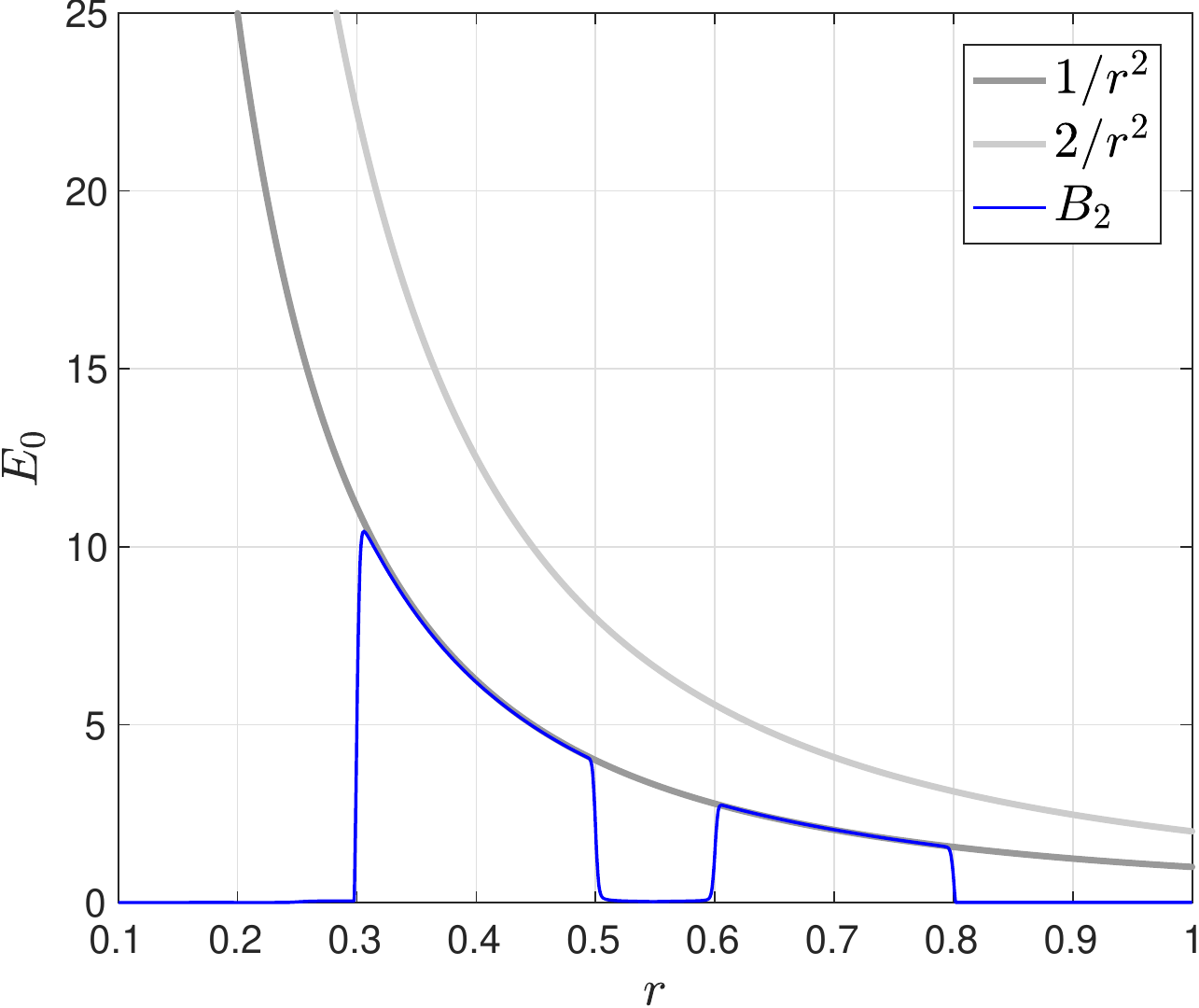}}
    \hfill
    \caption{Numerical results of the $B_2$ model of $E_0$ for bilateral beams}
  \label{fig:simpletest}
\end{figure}

\begin{figure}[htbp]
    \centering
    \subfigure[solution of $E_1$ at $t = 0$.]{  
    \includegraphics[width=0.3\textwidth,height=0.16\textheight]{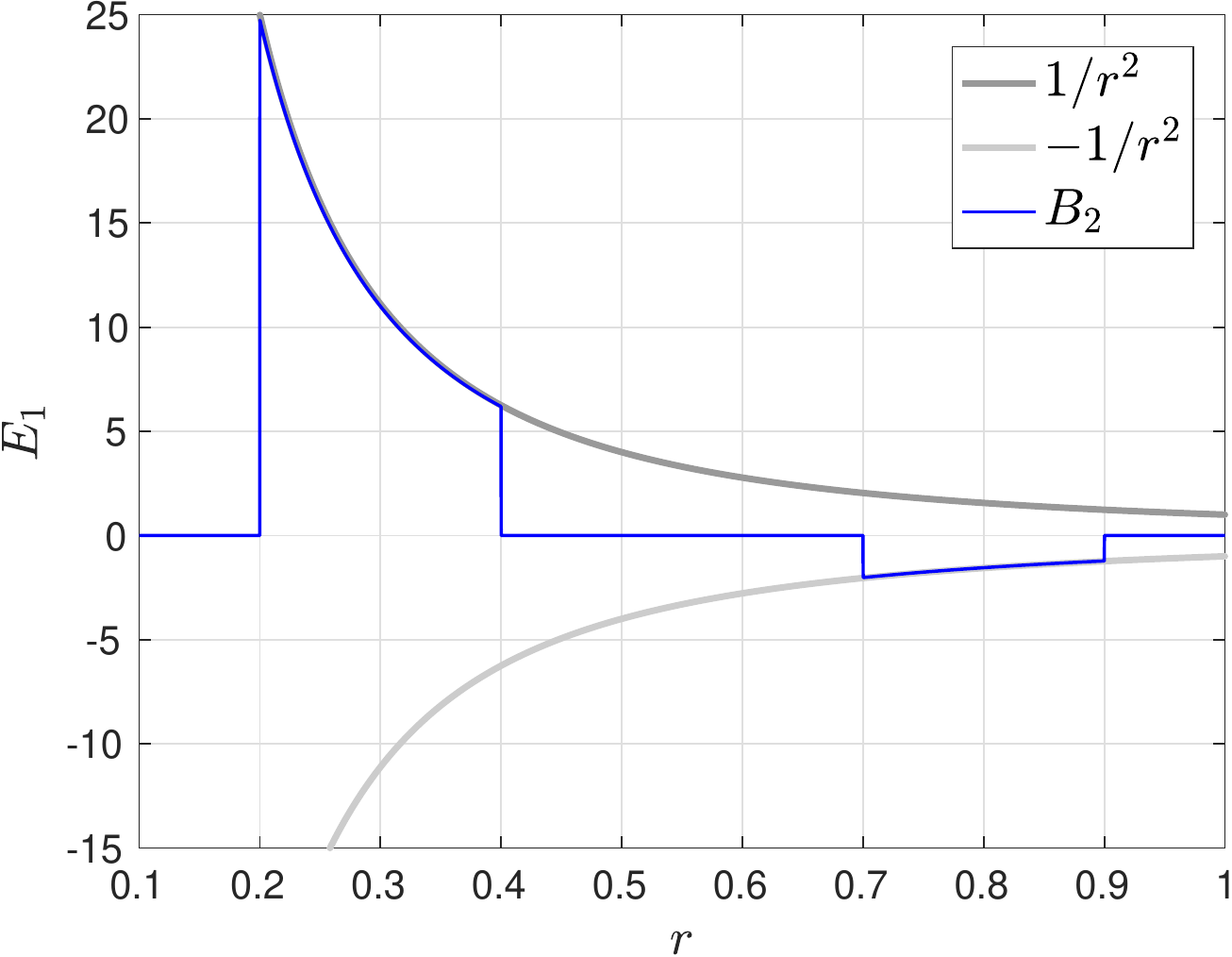}}
    \subfigure[solution of $E_1$ at $t = 0.1$.]{  
    \includegraphics[width=0.3\textwidth,height=0.16\textheight]{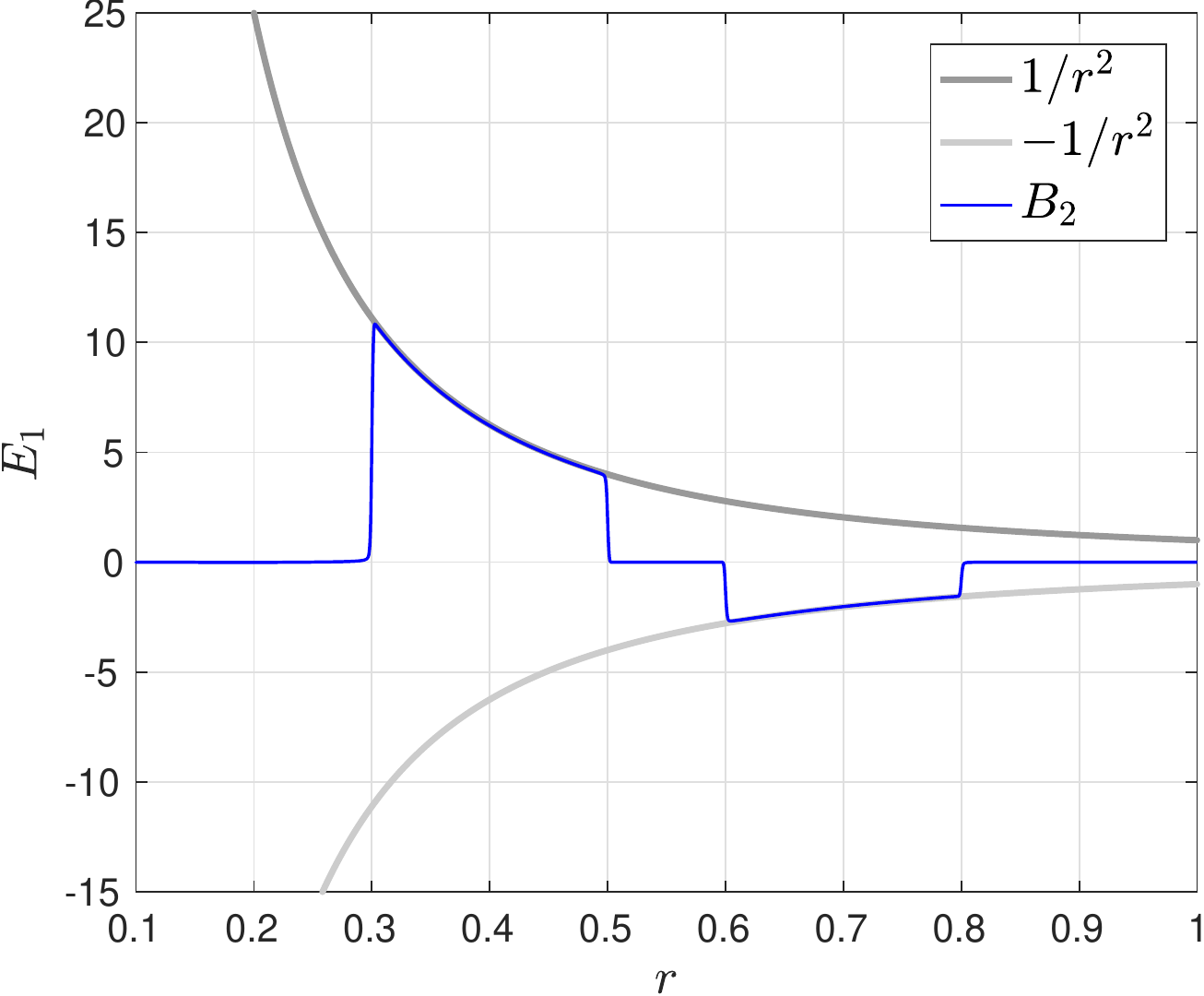}}
    \subfigure[solution of $E_1$ at $t = 0.2$.]{  
    \includegraphics[width=0.3\textwidth,height=0.16\textheight]{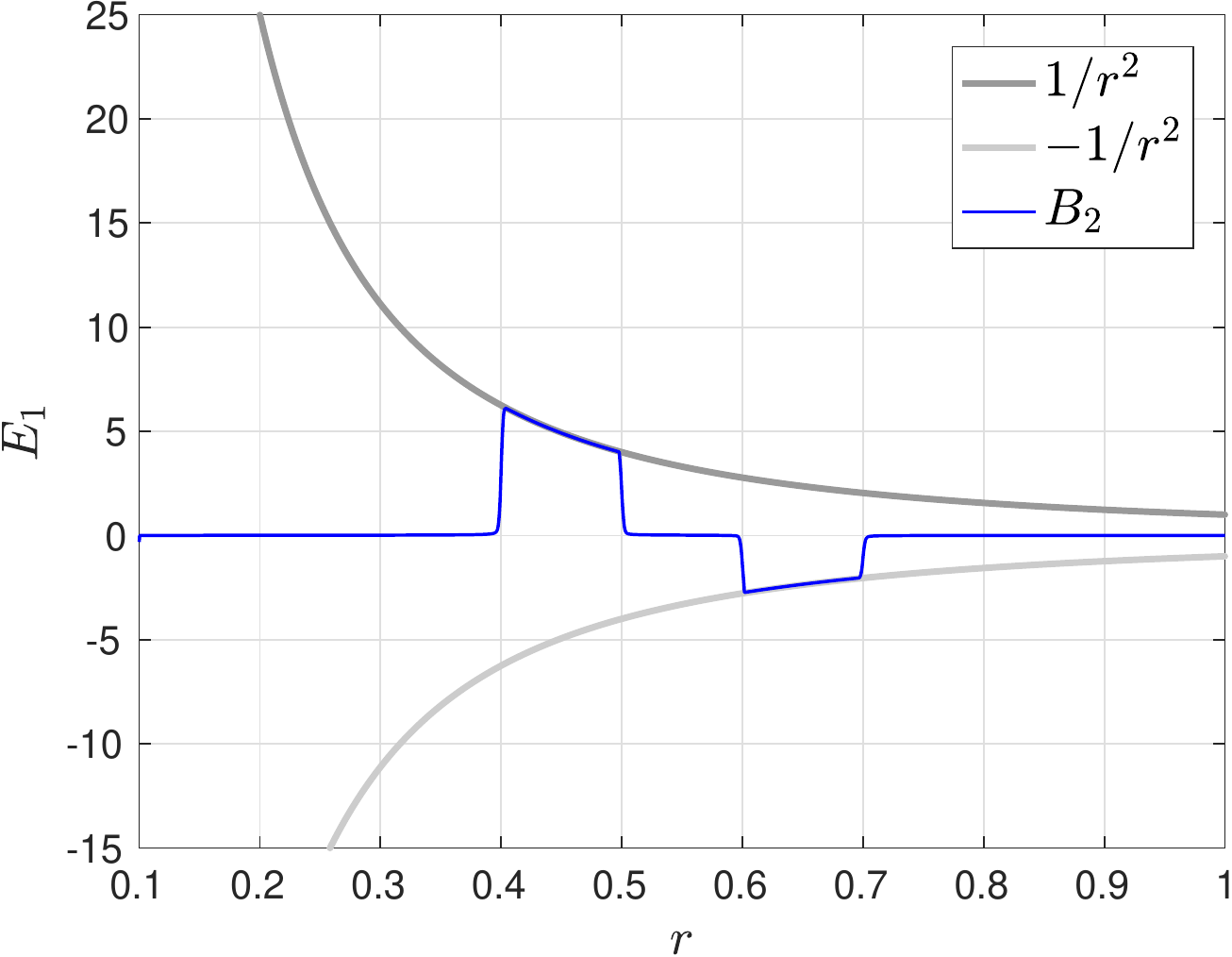}}
    \hfill
    \subfigure[solution of $E_1$ at $t = 0.25$.]{  
    \includegraphics[width=0.3\textwidth,height=0.16\textheight]{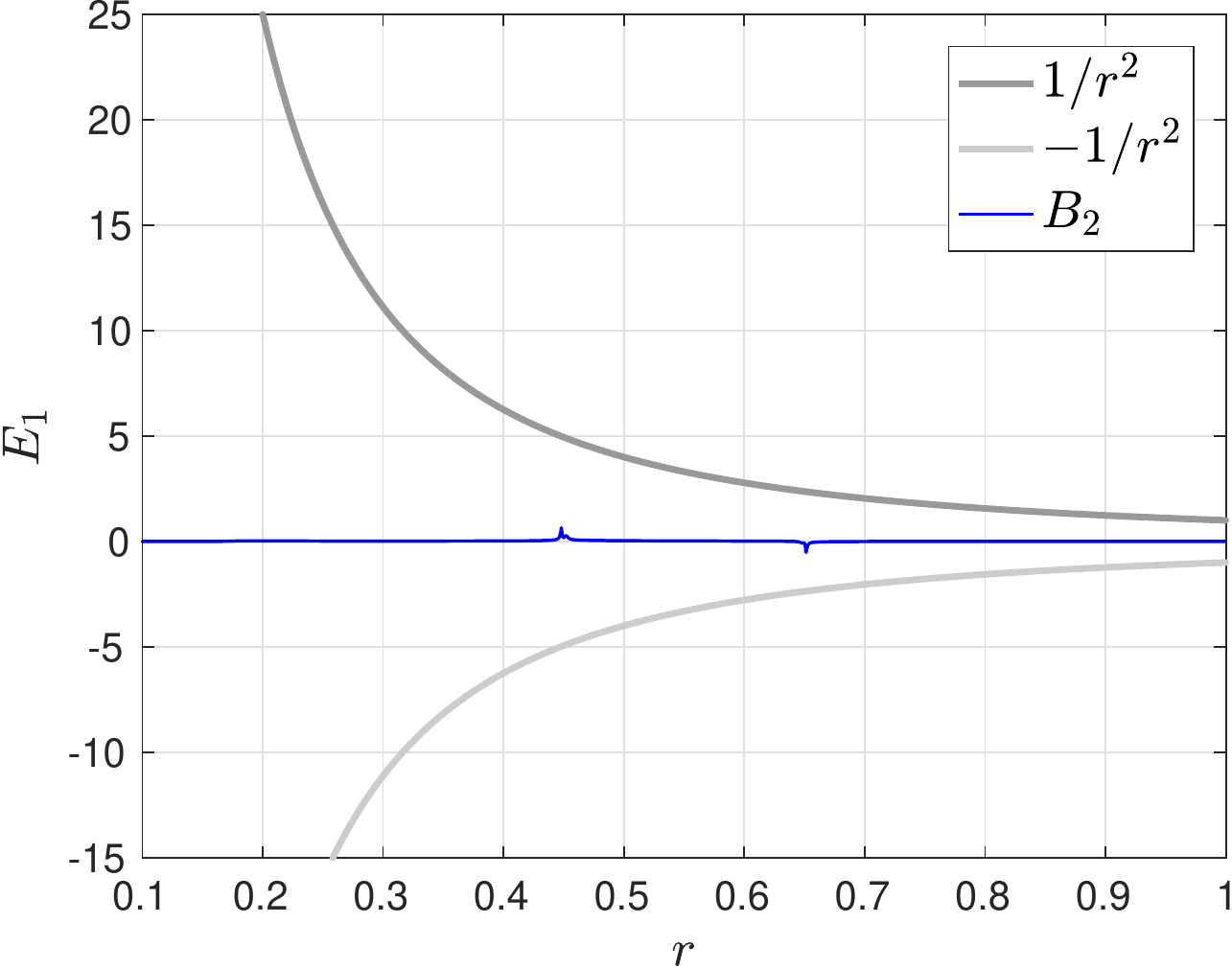}}
    \subfigure[solution of $E_1$ at $t = 0.3$.]{  
    \includegraphics[width=0.3\textwidth,height=0.16\textheight]{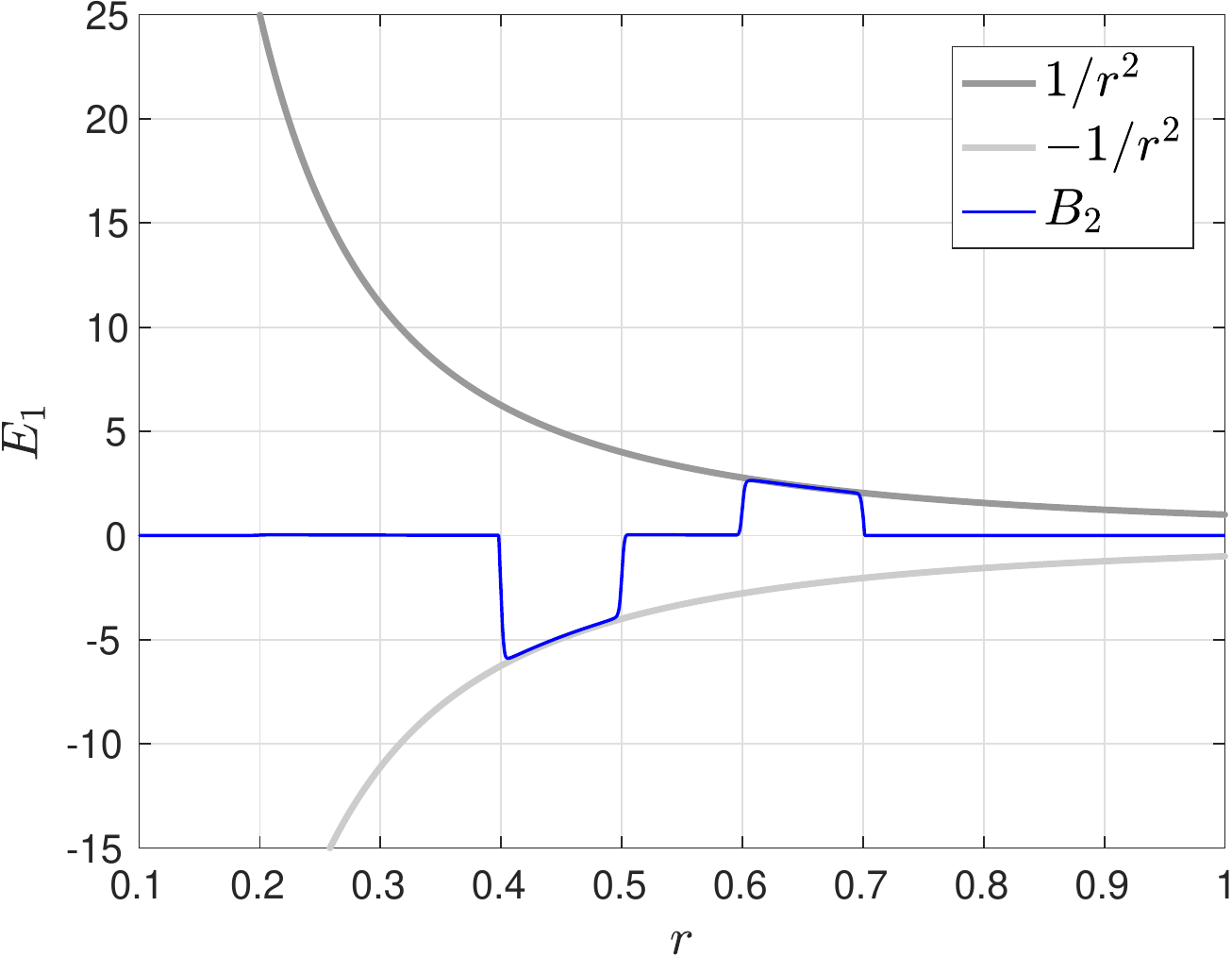}}
    \subfigure[solution of $E_1$ at $t = 0.4$.]{  
    \includegraphics[width=0.3\textwidth,height=0.16\textheight]{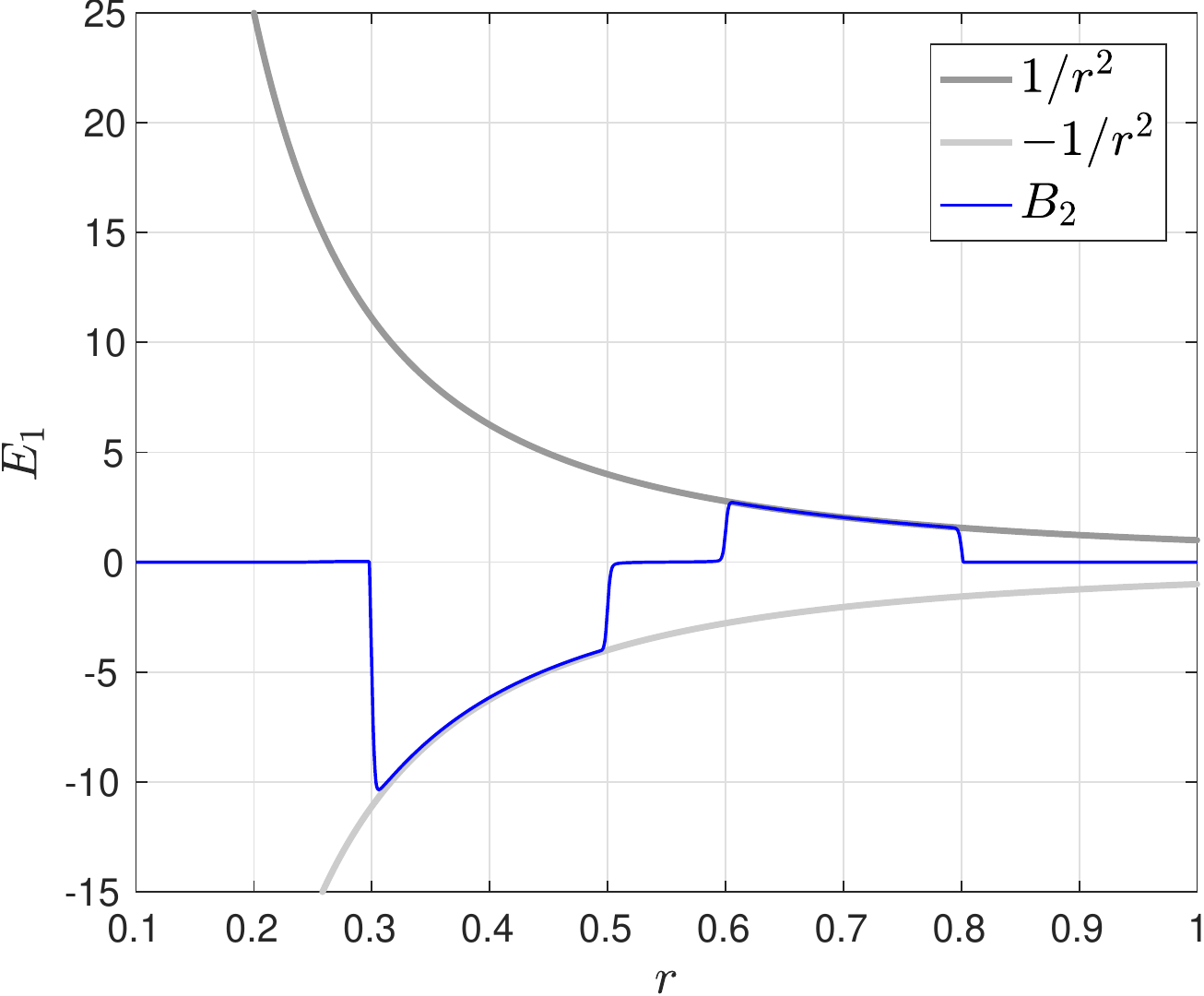}}
    \hfill
    \caption{Numerical results of the $B_2$ model of $E_1$ for bilateral beams }
  \label{fig:simpletestE1}
\end{figure}

\begin{figure}[htbp]
    \centering
    \subfigure[solution of $E_12$ at $t = 0$.]{  
    \includegraphics[width=0.3\textwidth,height=0.16\textheight]{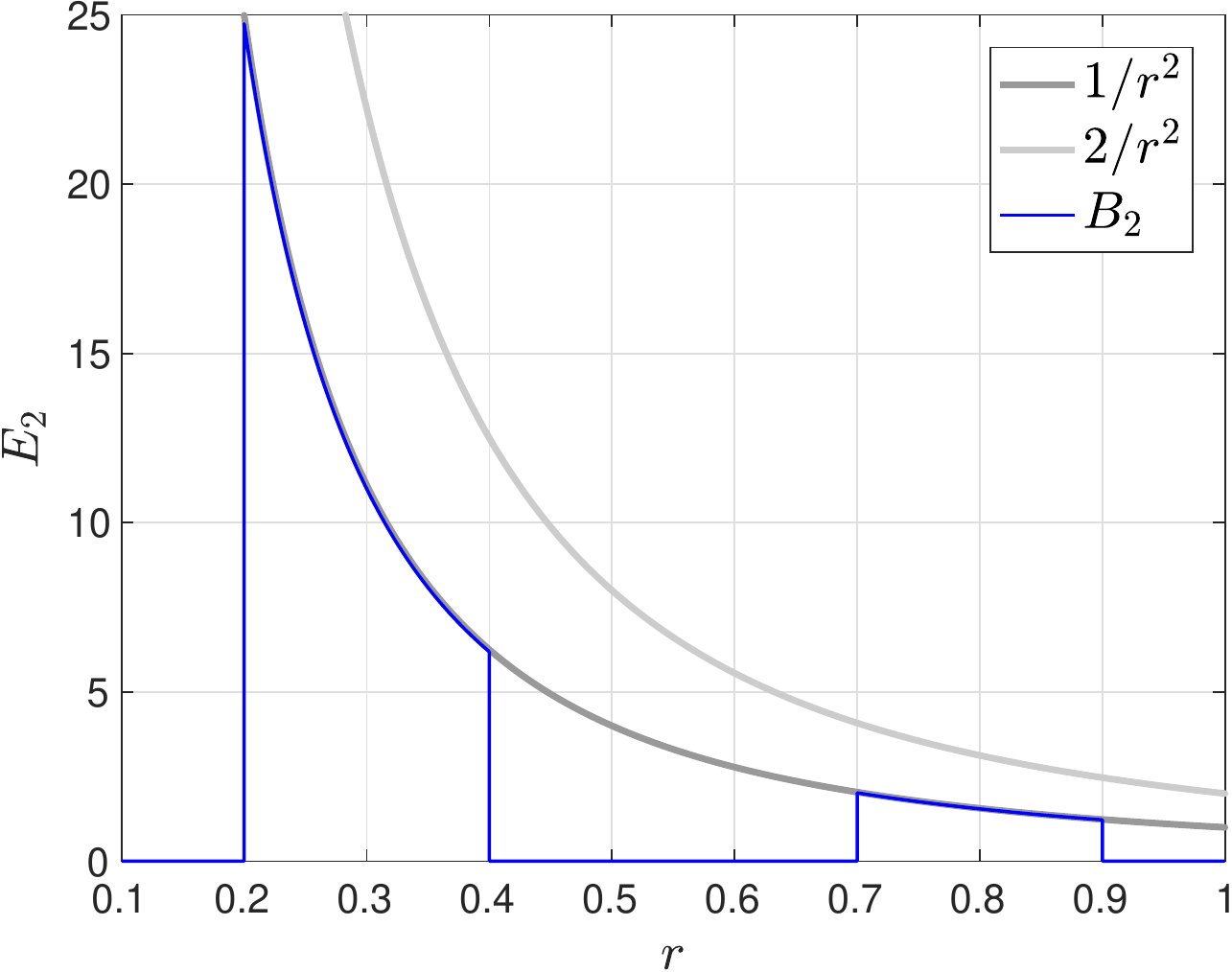}}
    \subfigure[solution of $E_2$ at $t = 0.1$.]{  
    \includegraphics[width=0.3\textwidth,height=0.16\textheight]{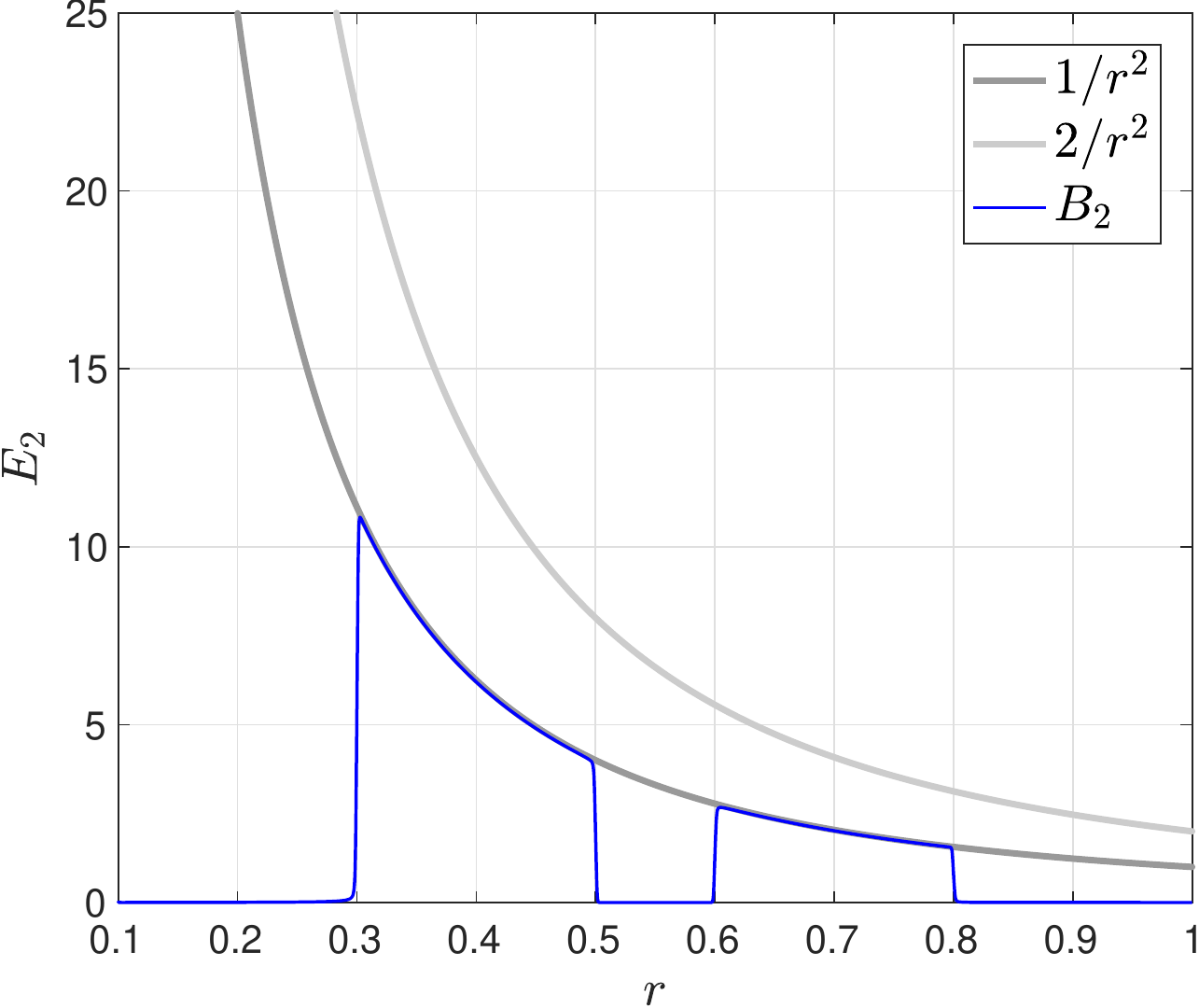}}
    \subfigure[solution of $E_2$ at $t = 0.2$.]{  
    \includegraphics[width=0.3\textwidth,height=0.16\textheight]{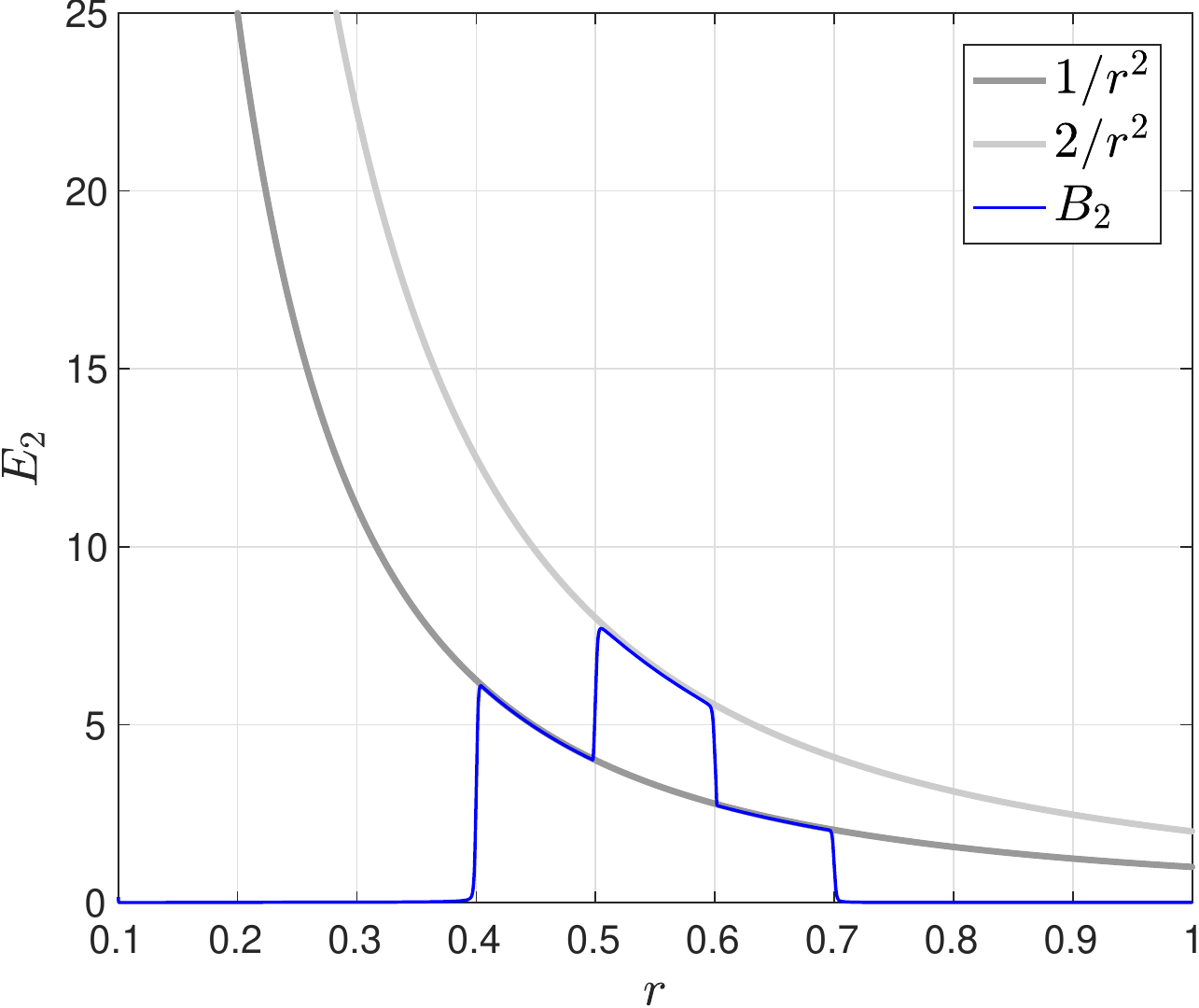}}
    \hfill
    \subfigure[solution of $E_2$ at $t = 0.25$.]{  
    \includegraphics[width=0.3\textwidth,height=0.16\textheight]{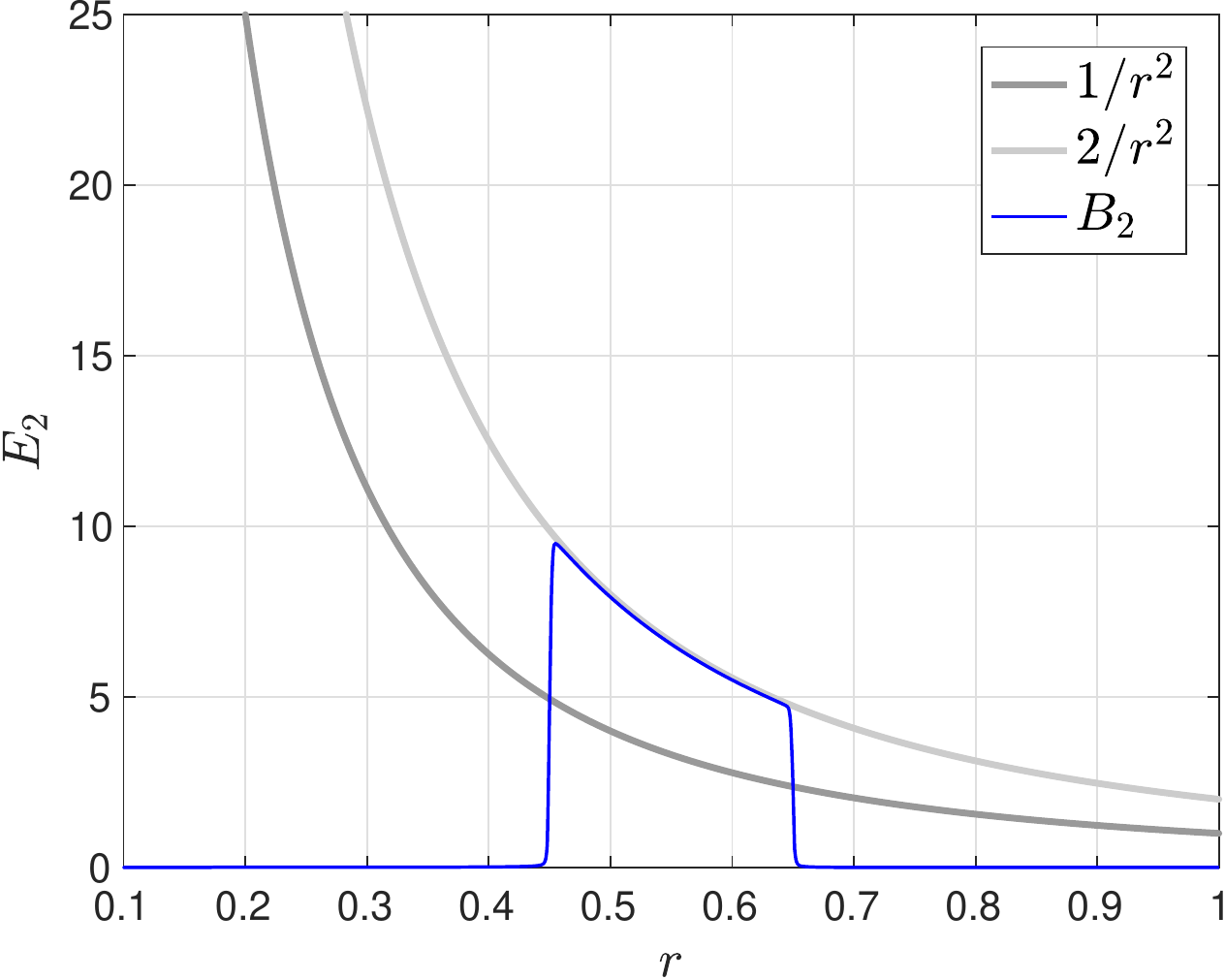}}
    \subfigure[solution of $E_2$ at $t = 0.3$.]{  
    \includegraphics[width=0.3\textwidth,height=0.16\textheight]{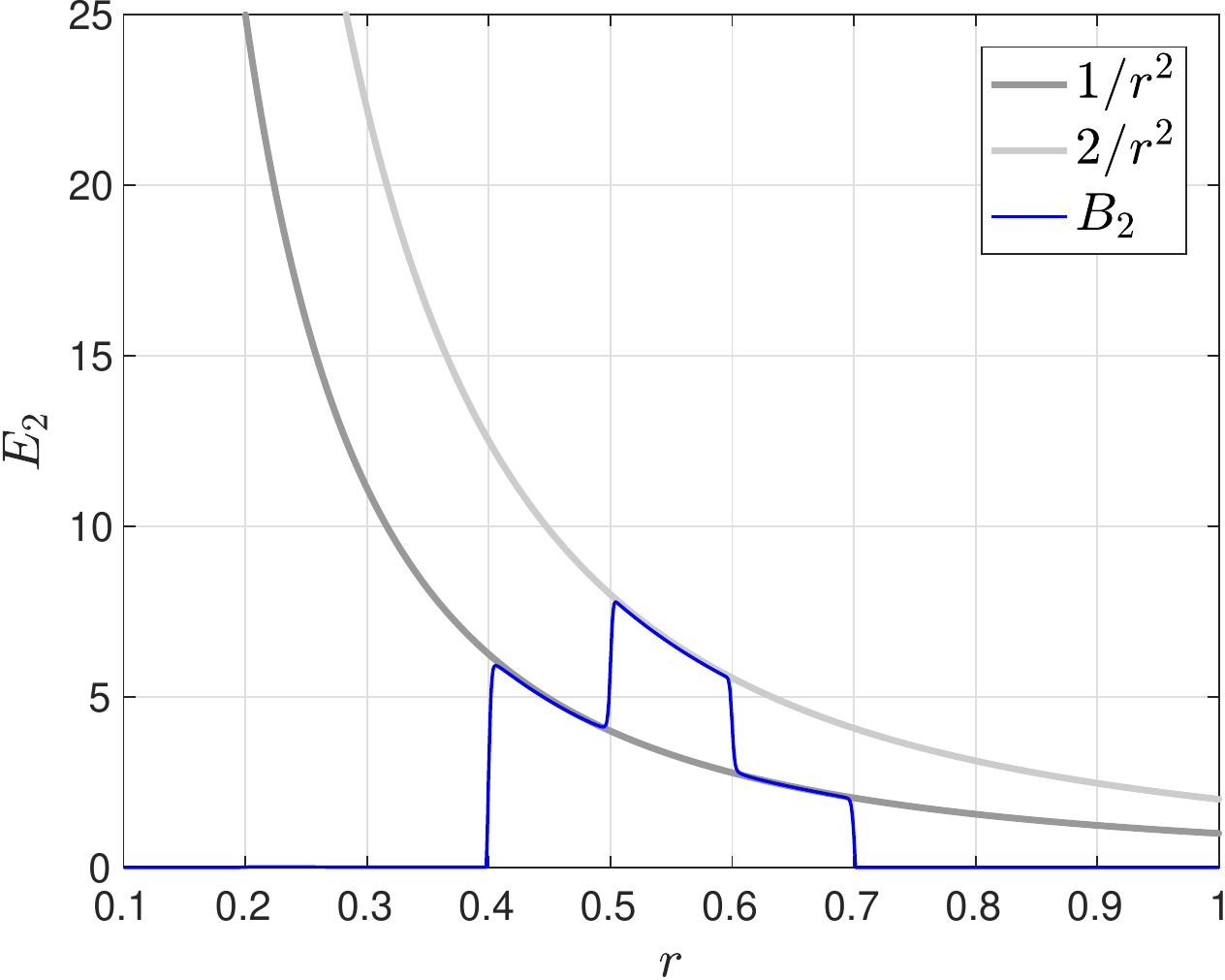}}
    \subfigure[solution of $E_2$ at $t = 0.4$.]{  
    \includegraphics[width=0.3\textwidth,height=0.16\textheight]{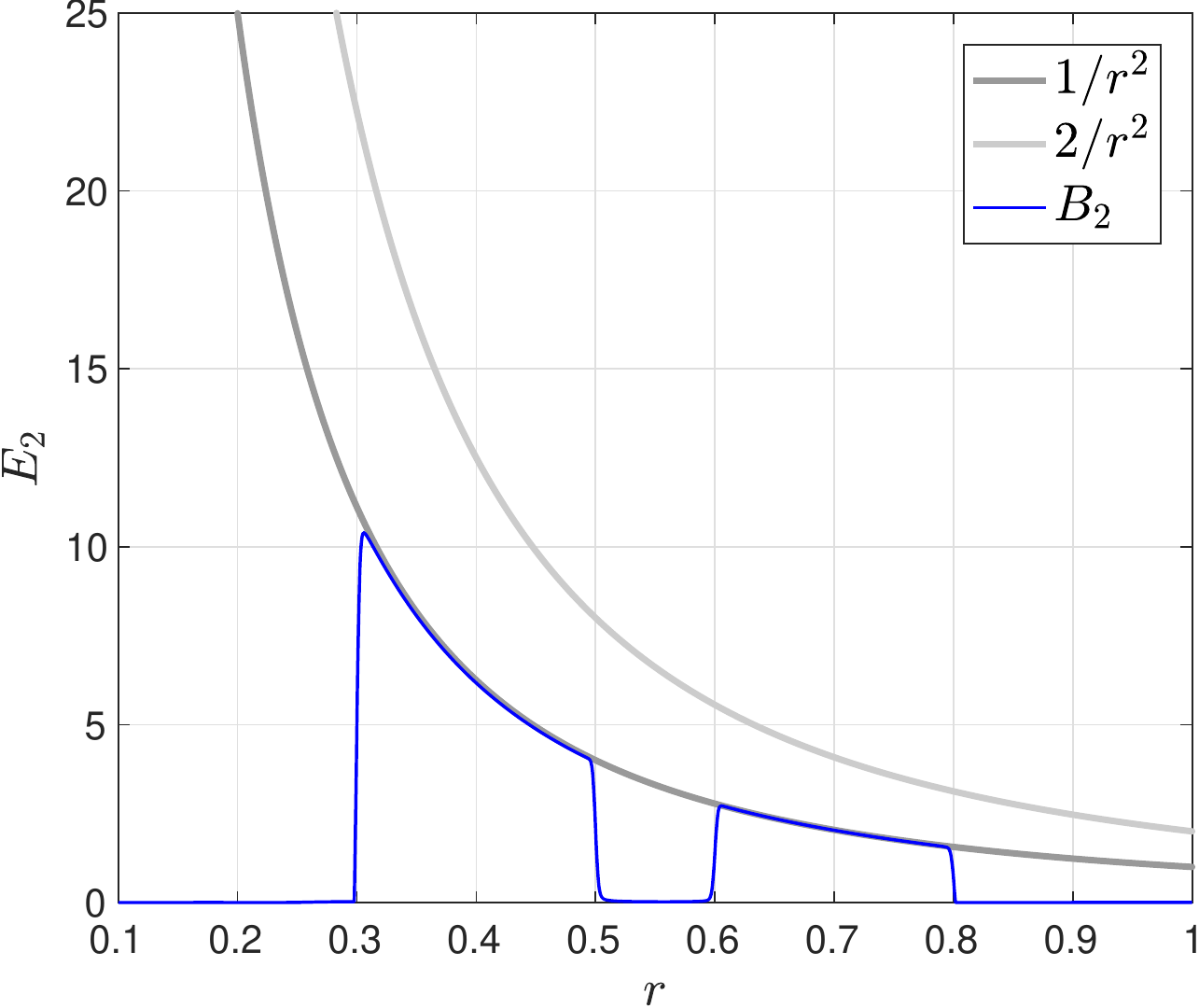}}
    \hfill
    \caption{Numerical results of the $B_2$ model of $E_2$ for the bilateral beams}
  \label{fig:simpletestE2}
\end{figure}

We simulate this problem with $N_{\text{cell}}=10000$ cells 
until $t_{end}=$0.1,0.2,0.25,0.3,0.4, and the
results of $E_0$, $E_1$, and $E_2$ are shown in \Cref{fig:simpletest},
\ref{fig:simpletestE1}, and \ref{fig:simpletestE2}, respectively.

\Cref{fig:simpletest} demonstrates that the $B_2$ model can obtain an
accurate solution of $E_0$ at all times. When $t=0$ and 0.1, the two waves are
independent; when $t=0.2, 0.25,$  and 0.3, the two waves intersect
each other; when $t=0.4$, the two waves separate.
Also, according to the results in \Cref{fig:simpletestE1} and
\Cref{fig:simpletestE2}, the $B_2$ solutions give first and second order
moments which are almost identical with the exact solution.
These results indicate that the $B_2$ model can describe this highly  
anisotropic specific intensity.

\begin{figure}[htbp]
    \centering
    \subfigure[Solution of $P_{10}$.]{  
    \includegraphics[width=0.3\textwidth,height=0.16\textheight]{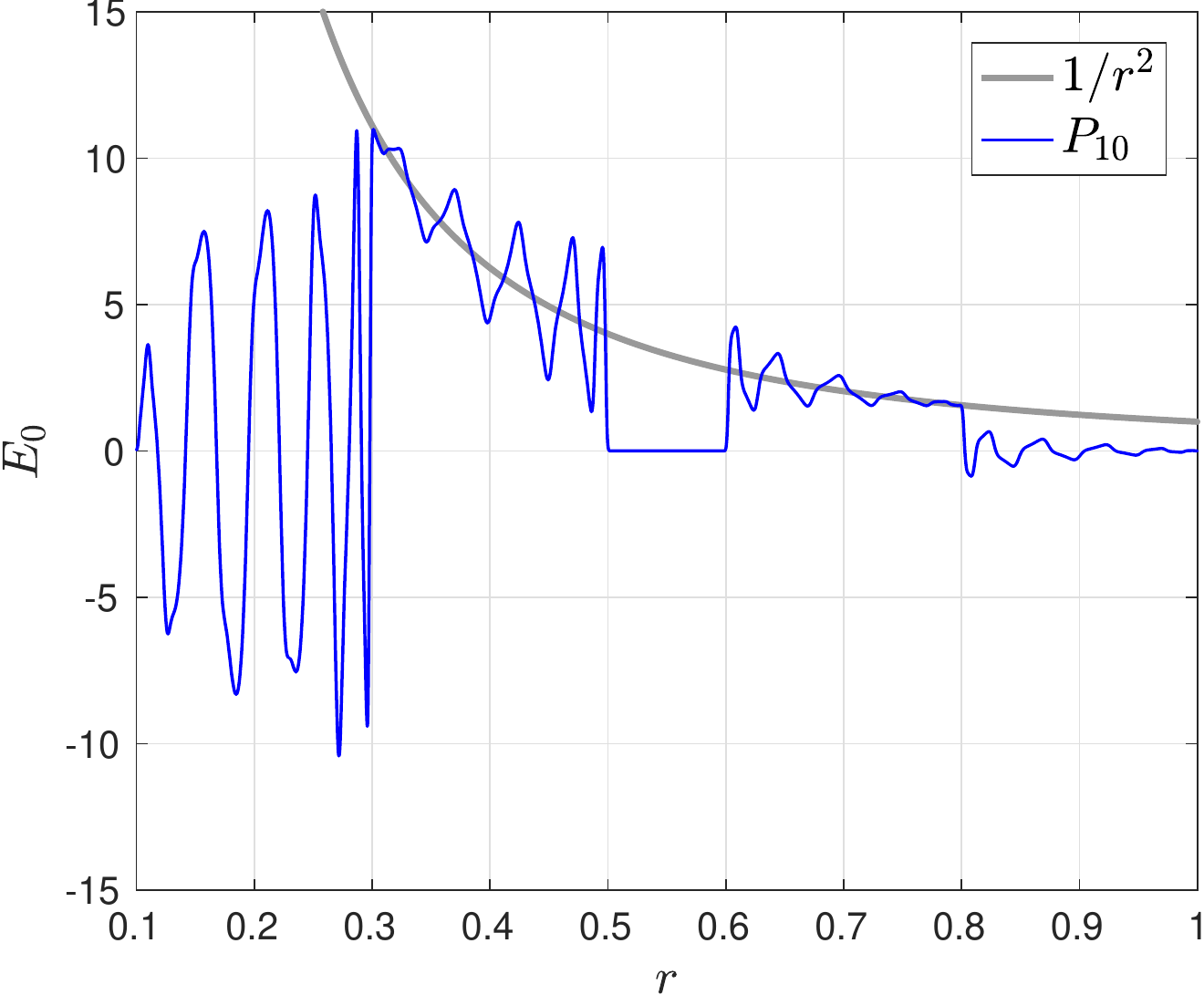}}
    \subfigure[Solution of $P_{30}$.]{  
    \includegraphics[width=0.3\textwidth,height=0.16\textheight]{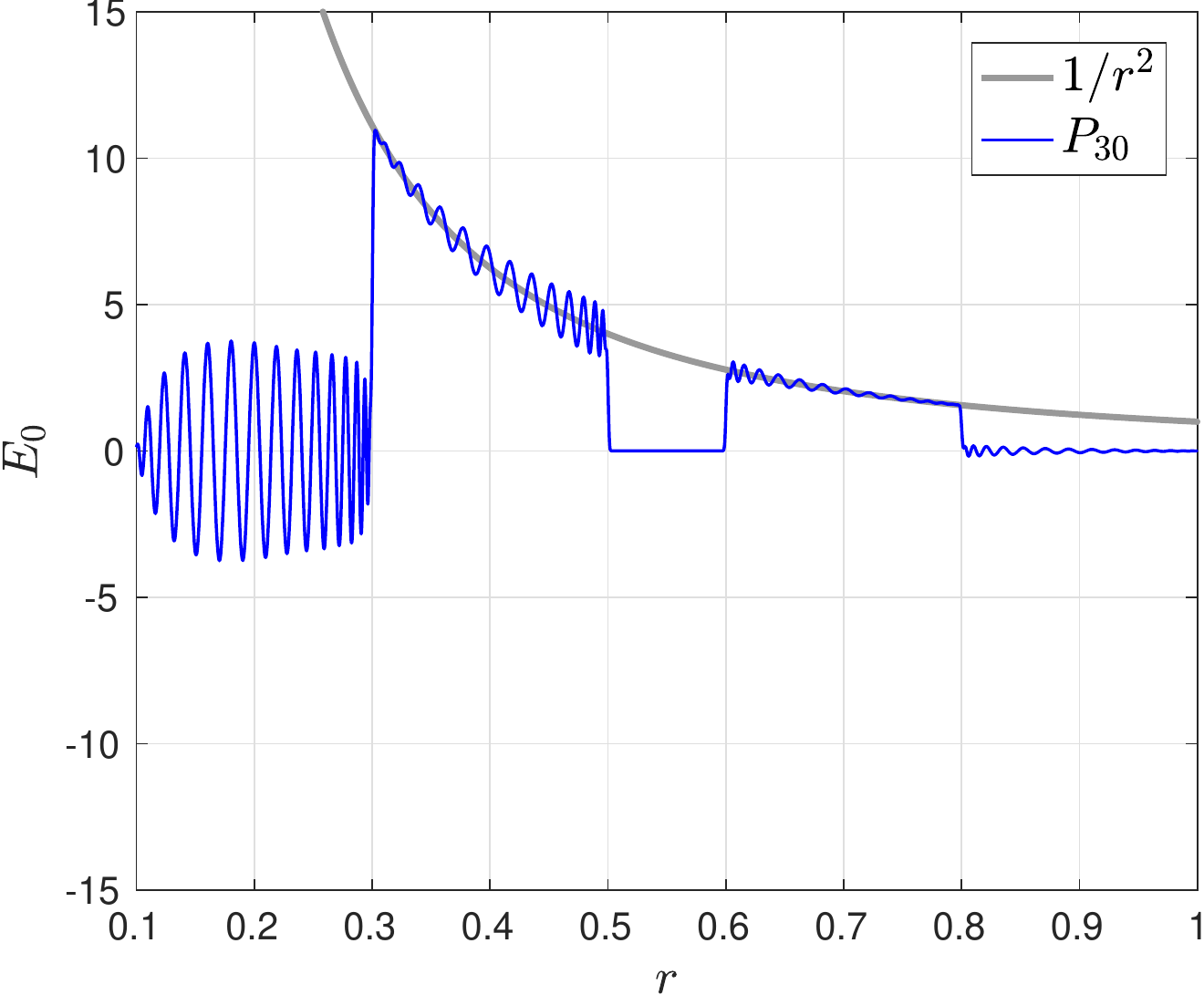}}
    \subfigure[Solution of $P_{70}$.]{  
    \includegraphics[width=0.3\textwidth,height=0.16\textheight]{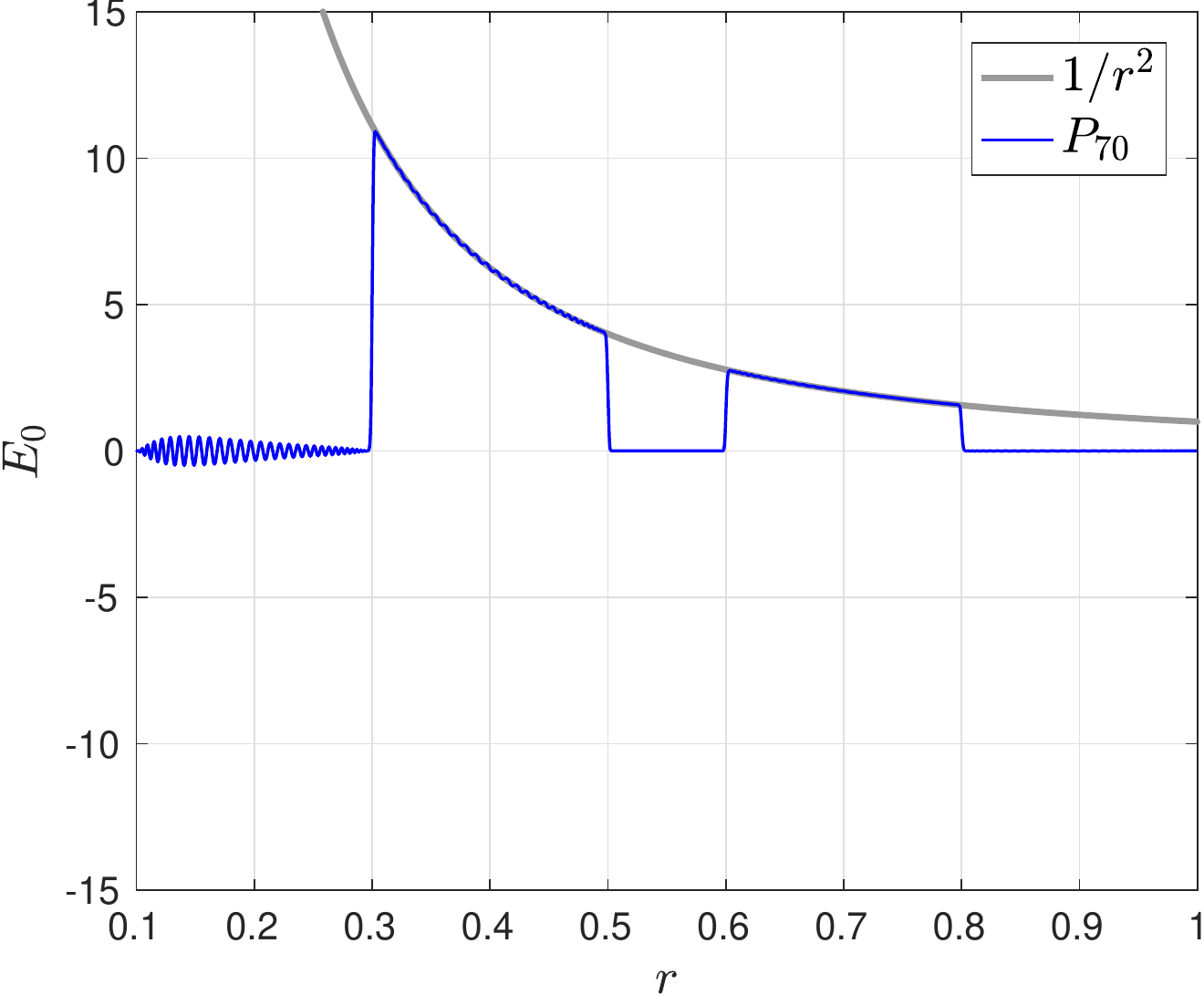}}
    \hfill
    \caption{Numerical results of the $P_N$ method for the bilateral
    beams}
  \label{fig:simpletestPn}
\end{figure}

We use the $P_N$ method with $N=$ 10, 30, and 70 to do the same
simulations, and the results of $E_0$ at $t=0.1$ are presented in
\Cref{fig:simpletestPn}. The $P_N$ solution of $N = 10$ and $30$ are
oscillatory and contains negative values for $E_0$, showing that the
$P_N$ method can not approximate this kind of beam-like distribution
well.  As $N$ gets larger, the oscillation decreases in magnitude.
However, the $P_N$ method requires to take as many as $N = 70$ to get
a sufficiently satisfying result.


\begin{example}[Laser beams propagating into vacuum]
  We use the same setup as \Cref{exm:example1}, where
  the opacities $\sigma_a=\sigma_s=0$, and the external source term
  $S=0$. The domain is unbounded (but $[0,10]$ for numerical
  experiments). The initial state is taken as  
\[
   I(0,r,\mu) = \left\{
   \begin{aligned}
     &\dfrac{1}{r^2}\delta(\mu-(r-5)), \quad &4\leq r\leq 6,\\
     &0,\quad & \text{otherwise}.
   \end{aligned}
   \right.
\]
\end{example}
The analytical solution can be obtained in the same way as in
\Cref{exm:example1}.  Direct calculation shows the exact solution of
this example to be
\[
   I(t,r,\mu) = 
   \left\{
   \begin{aligned}
    & \frac{1}{r^2(1+t)} \delta(\mu-\frac{r-5}{1+t}), \quad & 4-t\leq r \leq 6+t,\\
   &0, \quad &\text{otherwise}.
   \end{aligned}
   \right.
\]
In this problem, every point corresponds to a Dirac delta function,
which means that 
we need to approximate $\delta(\mu-\mu_0)$ for any 
$\mu_0\in[-1,1]$, because that at $r=\mu_0+5$, $4\leq \mu_0 \leq 6$,
the initial state is
$\dfrac{1}{r^2}\delta(\mu-\mu_0)$. This is difficult for commonly used 
numerical method, such as the discrete-ordinates ($S_N$) and the
spherical harmonics ($P_N$). 
However, consider the actual distribution function $I=\delta(\mu-\mu_0)$,
we have that 
\[ E_0=1,\quad E_1=\mu_0, \quad E_2 = \mu_0^2.\]
According to \eqref{eq:beta-closure}, we can obtain the moment closure
given by the $B_2$ model:
\[ 
  E_3 =\frac{E_1(E_2^2 + 2E_1^2 - 3E_0 E_2)}{2E_1^2 - E_0 E_2 - E_0^2}
  = \mu_0^3,
\]
which is the third moment of the actual distribution function.
Therefore, the $B_2$ model can approximate the Dirac delta function
$\delta(\mu-\mu_0)$ for any $\mu\in[-1,1]$.

\begin{figure}[htbp]
    \centering
    \subfigure[Solution of $E_0$]{  
    \includegraphics[width=0.3\textwidth,height=0.16\textheight]{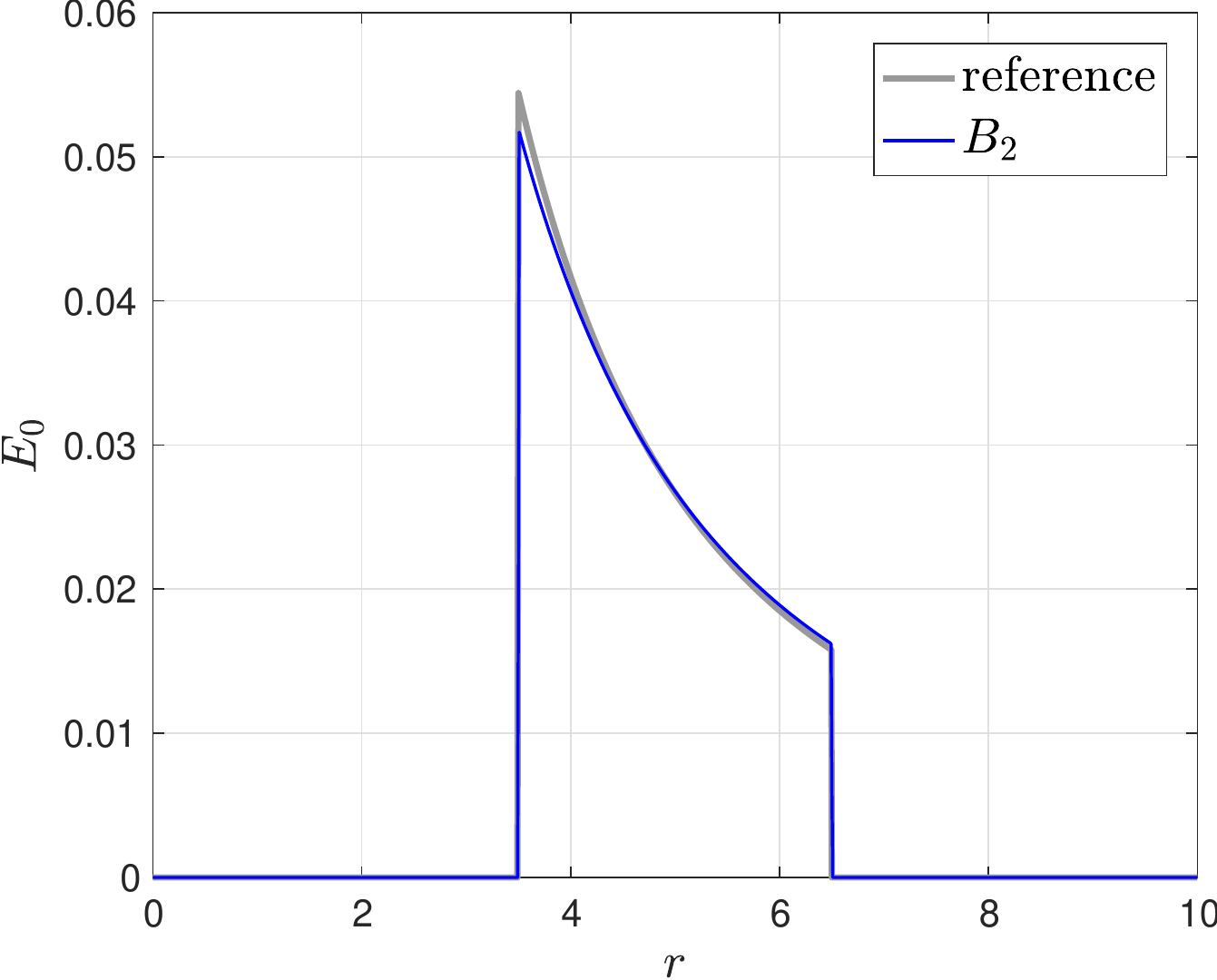}}
    \subfigure[Solution of $\dfrac{E_1}{E_0}$]{  
    \includegraphics[width=0.3\textwidth,height=0.16\textheight]{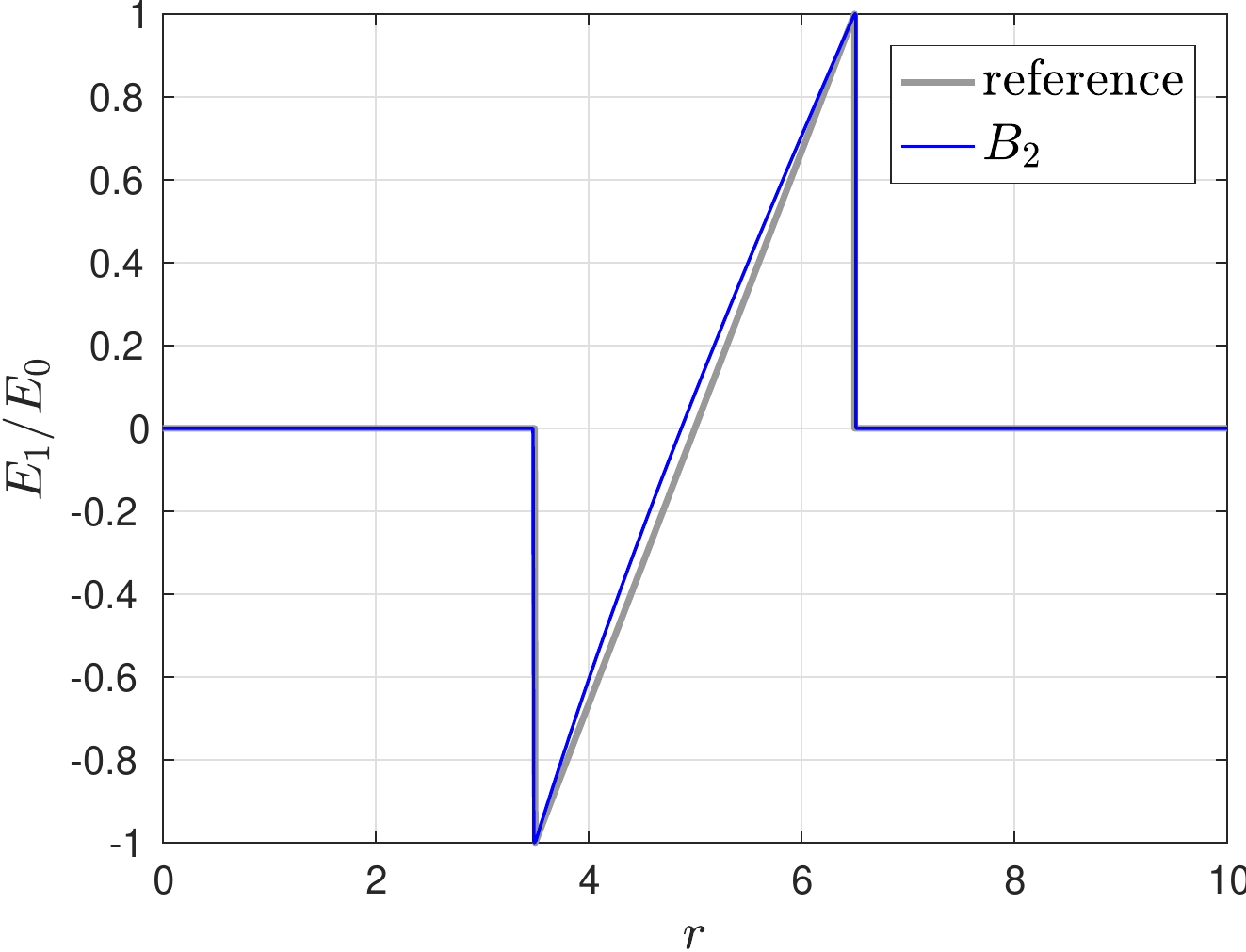}}
    \subfigure[Solution of $\dfrac{E_2}{E_0}$]{  
    \includegraphics[width=0.3\textwidth,height=0.16\textheight]{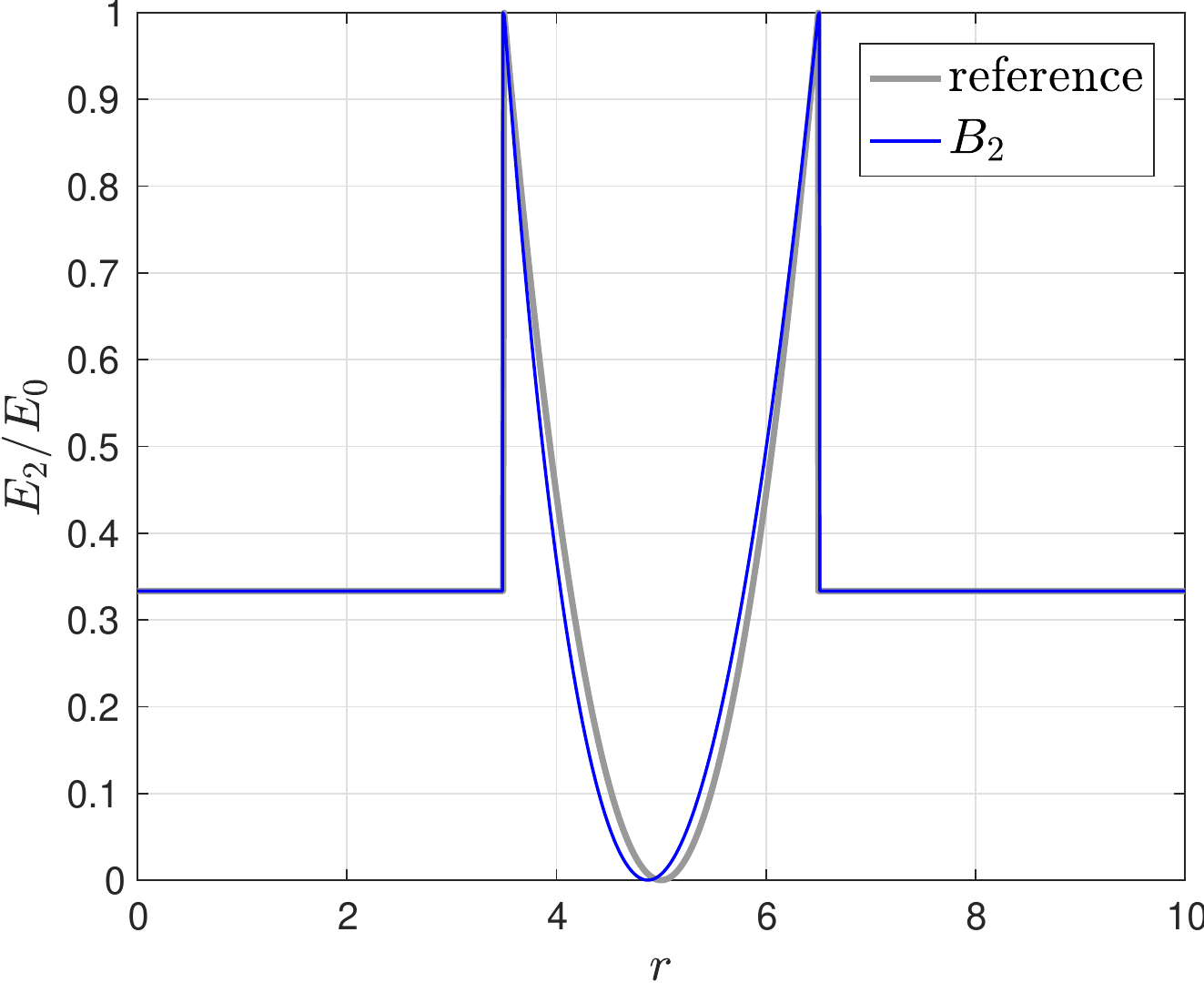}}
    \caption{Numerical results of the $B_2$ method for laser beams
    propagating into vacuum}
  \label{fig:simpletest2}
\end{figure}

\begin{figure}[htbp]
    \centering
    \subfigure[Solution of $P_{10}$]{  
    \includegraphics[width=0.3\textwidth,height=0.16\textheight]{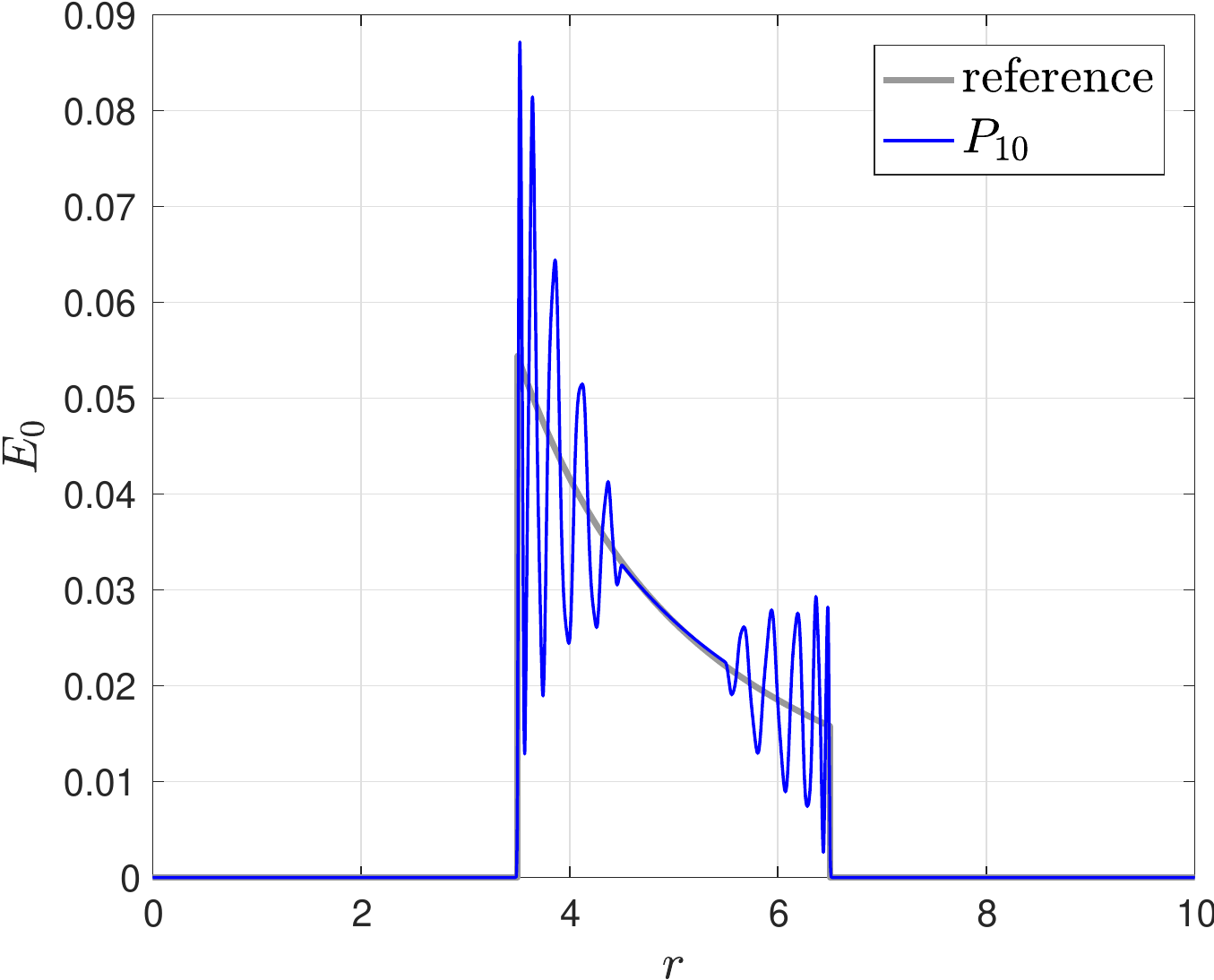}}
    \subfigure[Solution of $P_{30}$]{  
    \includegraphics[width=0.3\textwidth,height=0.16\textheight]{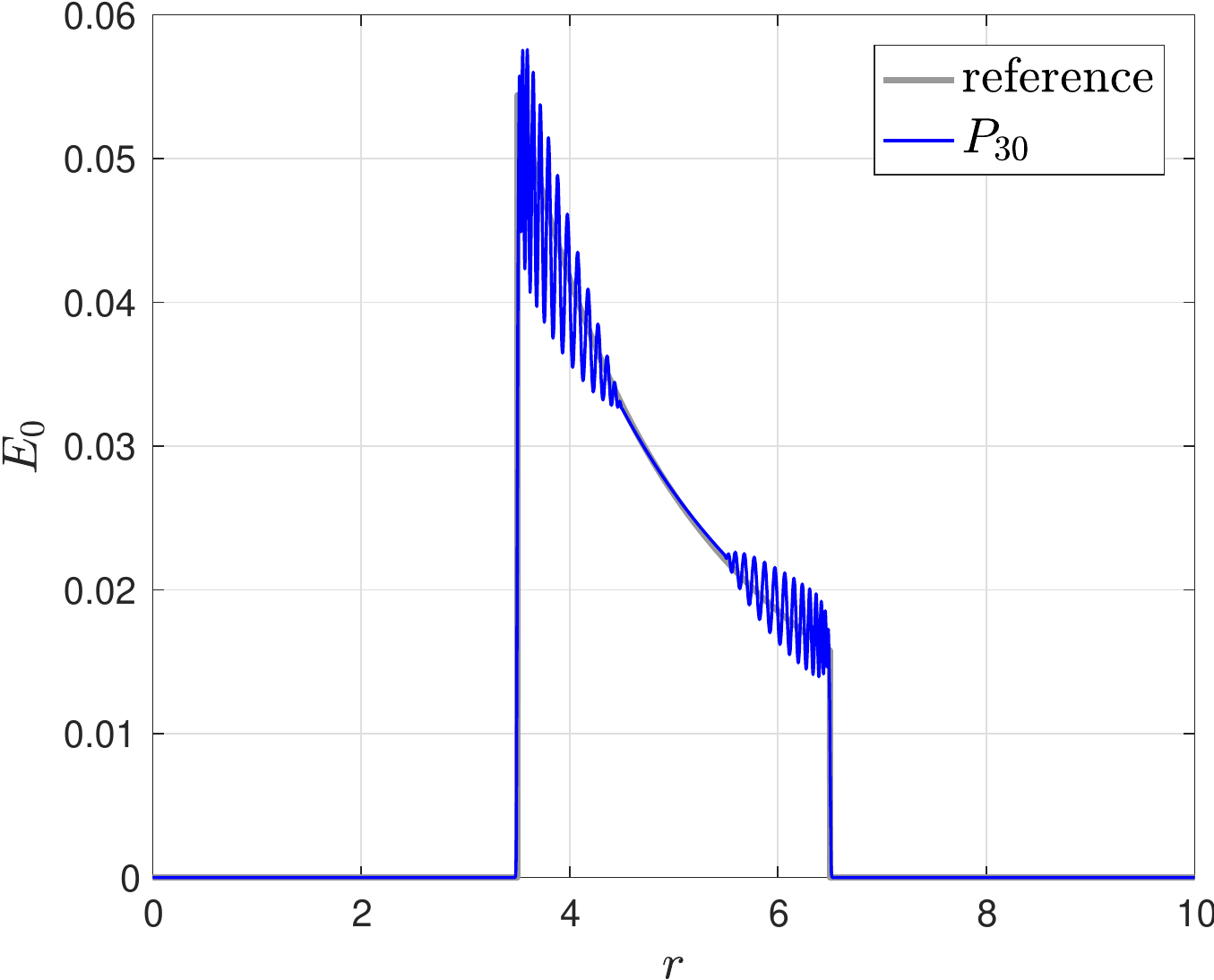}}
    \subfigure[Solution of $P_{50}$]{  
    \includegraphics[width=0.3\textwidth,height=0.16\textheight]{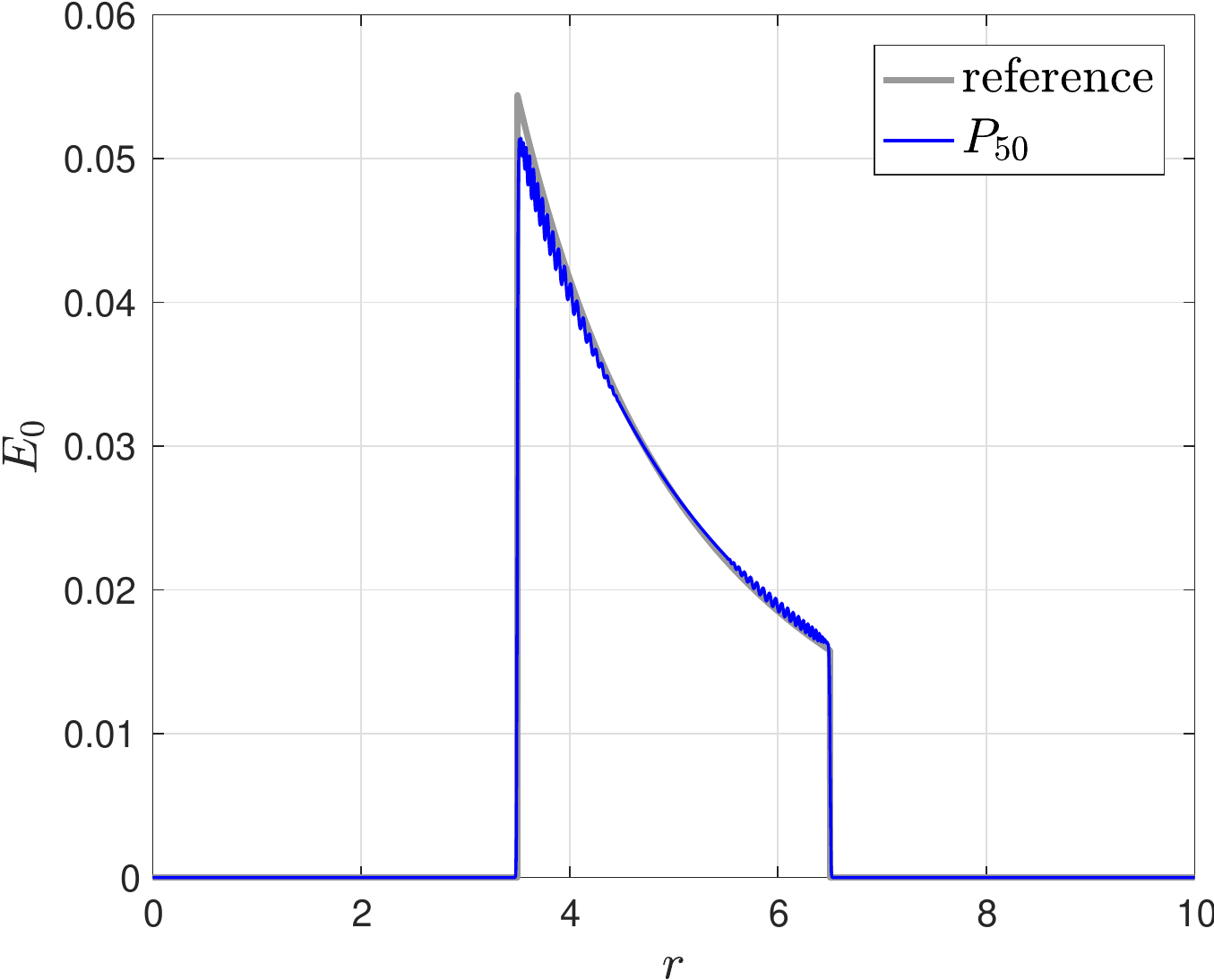}}
    \caption{Numerical results of the $P_N$ method for laser beams
    propagating into vacuum}
  \label{fig:simpletest2Pn}
\end{figure}

We use $N_{\text{cell}}=100000$ cells to simulate
this problem until $t=0.5$, and the results of $E_0$, $E_1$, and $E_2$
are shown in \Cref{fig:simpletest2}.
Results in \Cref{fig:simpletest2} show that the solutions of all the
three moments considered in the $B_2$ model agree with the reference
solution, demonstrating the effectiveness of the 
$B_2$ model in approximating this test case where the distribution
function contain the Dirac function $\delta(\mu-\mu_0)$.
Moreover, we use the $P_N$ method with $N=$ 10, 30, and 50 
to do the same simulations, and the
results of $E_0$ are shown in \Cref{fig:simpletest2Pn}. 
There are many
oscillations in the results of the $P_N$ model, and as many as $N =
50$ are required in order to get a satisfying result.
Therefore, the $P_N$ model is not effective for describing 
this kind of strongly anisotropic distribution.

\begin{example}[Homogeneous sphere]
This example consists of a static homogeneous and isothermal sphere of
radius $R$ that radiates into vacuum. The radiation interacts with the
background matter only via isotropic absorption and thermal emission
inside the sphere. 
\begin{equation}
  \mu \pd{I}{r} + \dfrac{1-\mu^2}{r}\pd{I}{\mu} = 
  \left\{
  \begin{aligned}
  & \sigma_a(B-I),\quad &r\leq R, \\
  & 0, \quad &r>R.
  \end{aligned}
  \right.
\end{equation}
The sphere radius is $R=1$ and the spatial domain is unbounded (but
$[0,5]$ for numerical experiments). Inside the sphere the absorption
and scattering opacities are chosen to be $\sigma_a=250$ and
$\sigma_s=0$, respectively, and the emissivity $B=1$; while in the
vacuum at $r>R$, we have $\sigma_a=\sigma_s=B=0$.  The boundary
condition at $r=5$ is an infinite boundary condition, and the initial
state is $I(\mu)=0.5\times 10^{-7}$ everywhere.
\end{example}
As this test case contains a sharp transition from the diffusive
regime inside the sphere to the free-streaming regime outside, it is
frequently employed to test radiative transfer codes
\cite{Schinder1989Radiative,Smit1997Hyperbolicity,abdikamalov2012new}.

The steady-state transport equation can be solved
analytically \cite{Smit1997Hyperbolicity}:
\begin{align}
  I(r,\mu) = B(1-e^{-\sigma_as(r,\mu)}),
\end{align}
where 
\begin{align}
  s(r,\mu) = \left\{
  \begin{aligned}
    &r\mu + Rg(r,\mu)\quad &\text{if   } r\leq R, \quad 
    &-1\leq \mu\leq 1,\\
    &2Rg(r,\mu)\quad &\text{if   } r>R, \quad 
    &\sqrt{1-\left( \frac{R}{r} \right)^2}\leq \mu\leq 1,\\
    &0,\quad &\text{otherwise} 
  \end{aligned}
  \right.
\end{align}
and 
\begin{align}
  g(r,\mu) = \sqrt{1-\left(\frac{r}{R}\right)^2(1-\mu^2)}.
\end{align}

\begin{figure}[htbp]
    \centering
    \subfigure[Solution of $E_0$.]{  
    \includegraphics[width=0.3\textwidth,height=0.16\textheight]{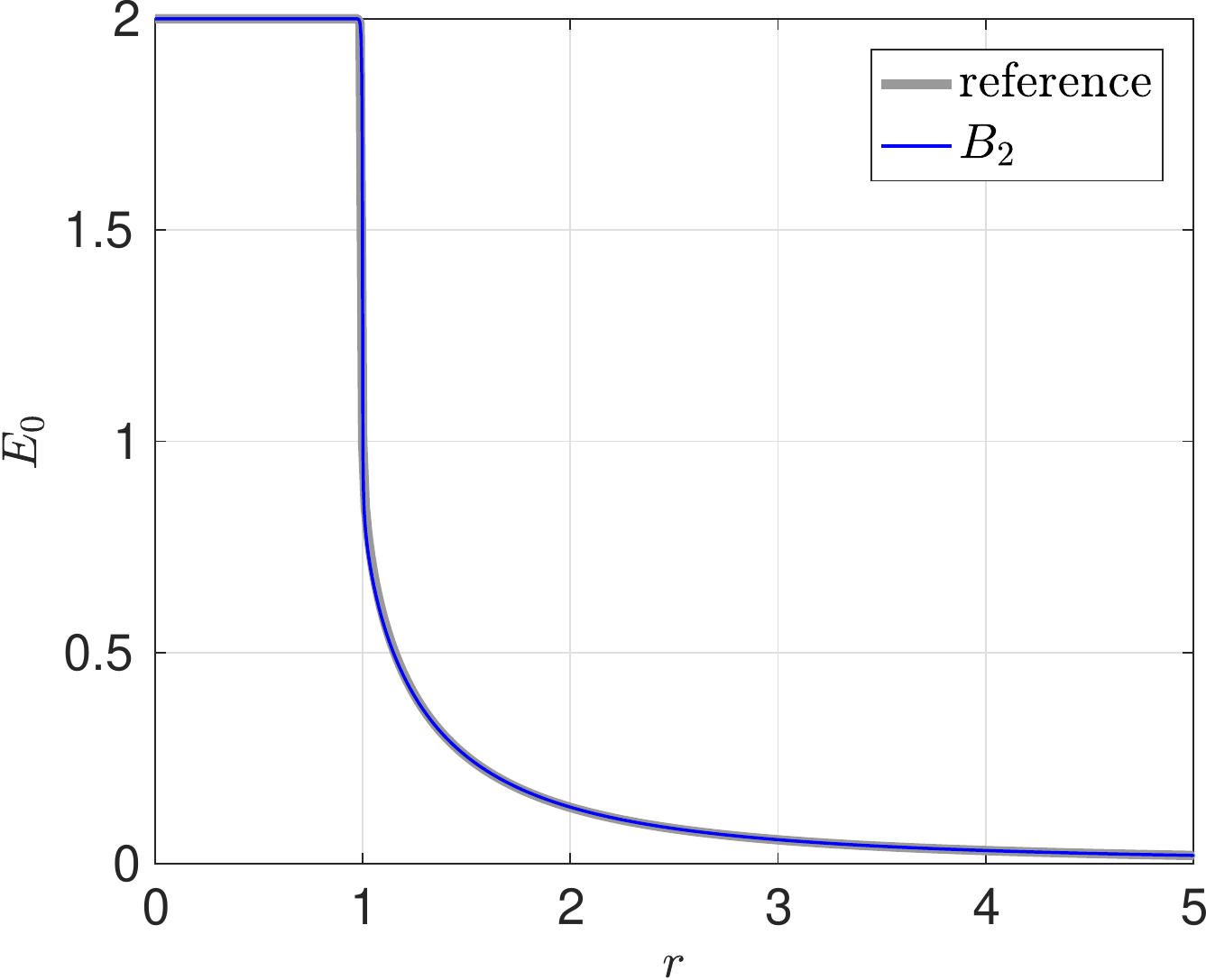}}
    \subfigure[Solution of $E_1$.]{
    \includegraphics[width=0.3\textwidth,height=0.16\textheight]{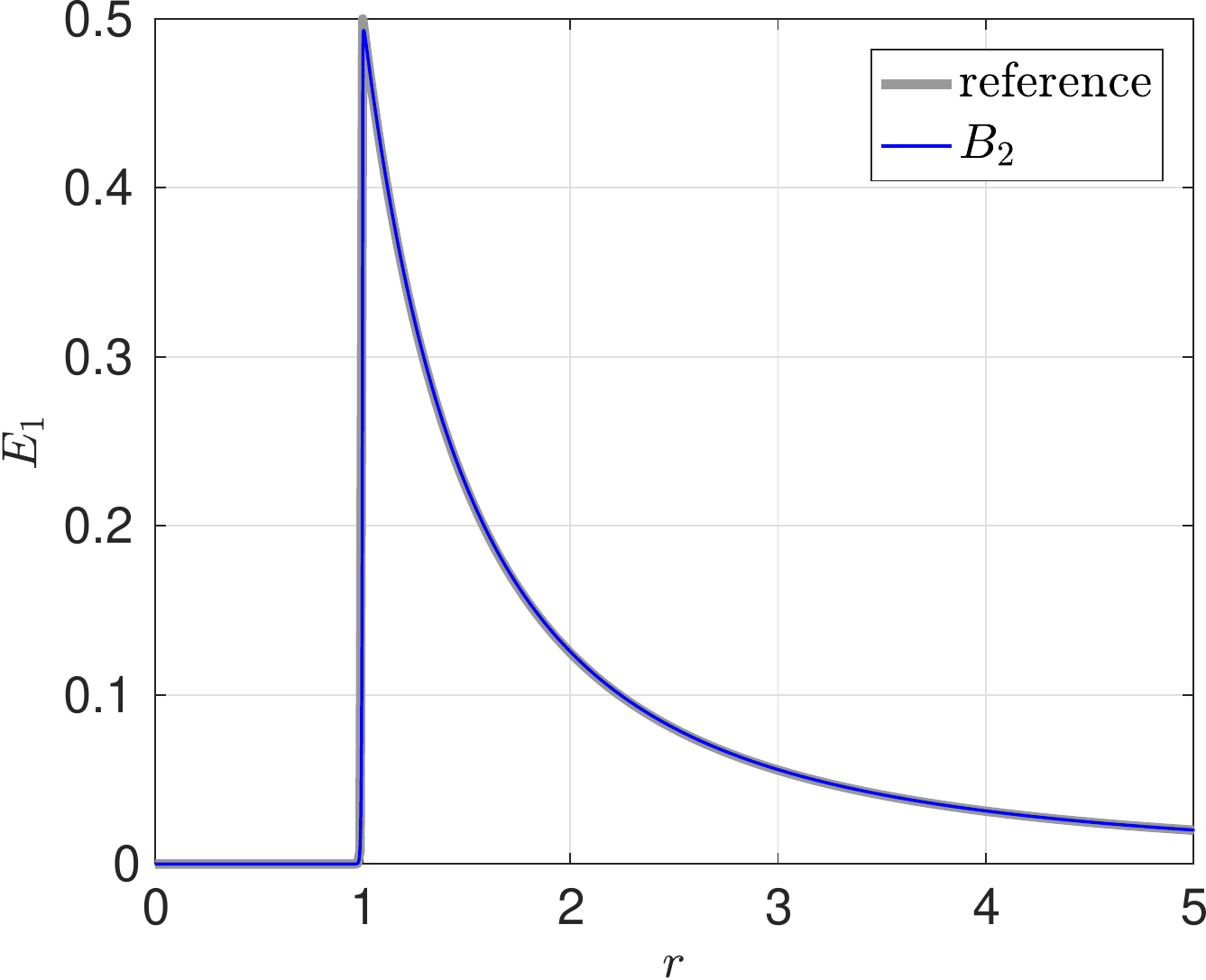}}
    \subfigure[Solution of $E_2$.]{
    \includegraphics[width=0.3\textwidth,height=0.16\textheight]{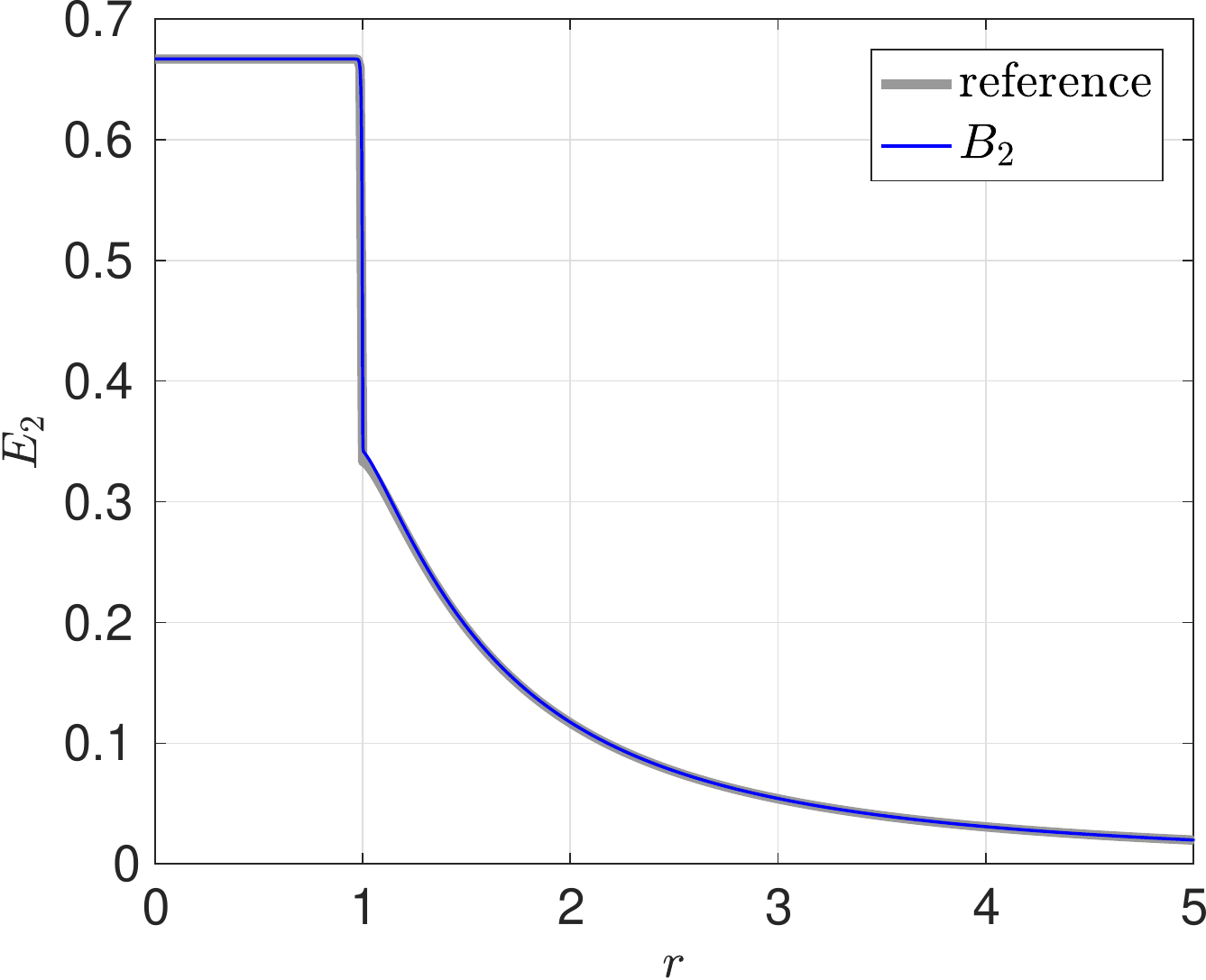}}
    \hfill
    \subfigure[Solution of $E_1/E_0$.]{
    \includegraphics[width=0.3\textwidth,height=0.16\textheight]{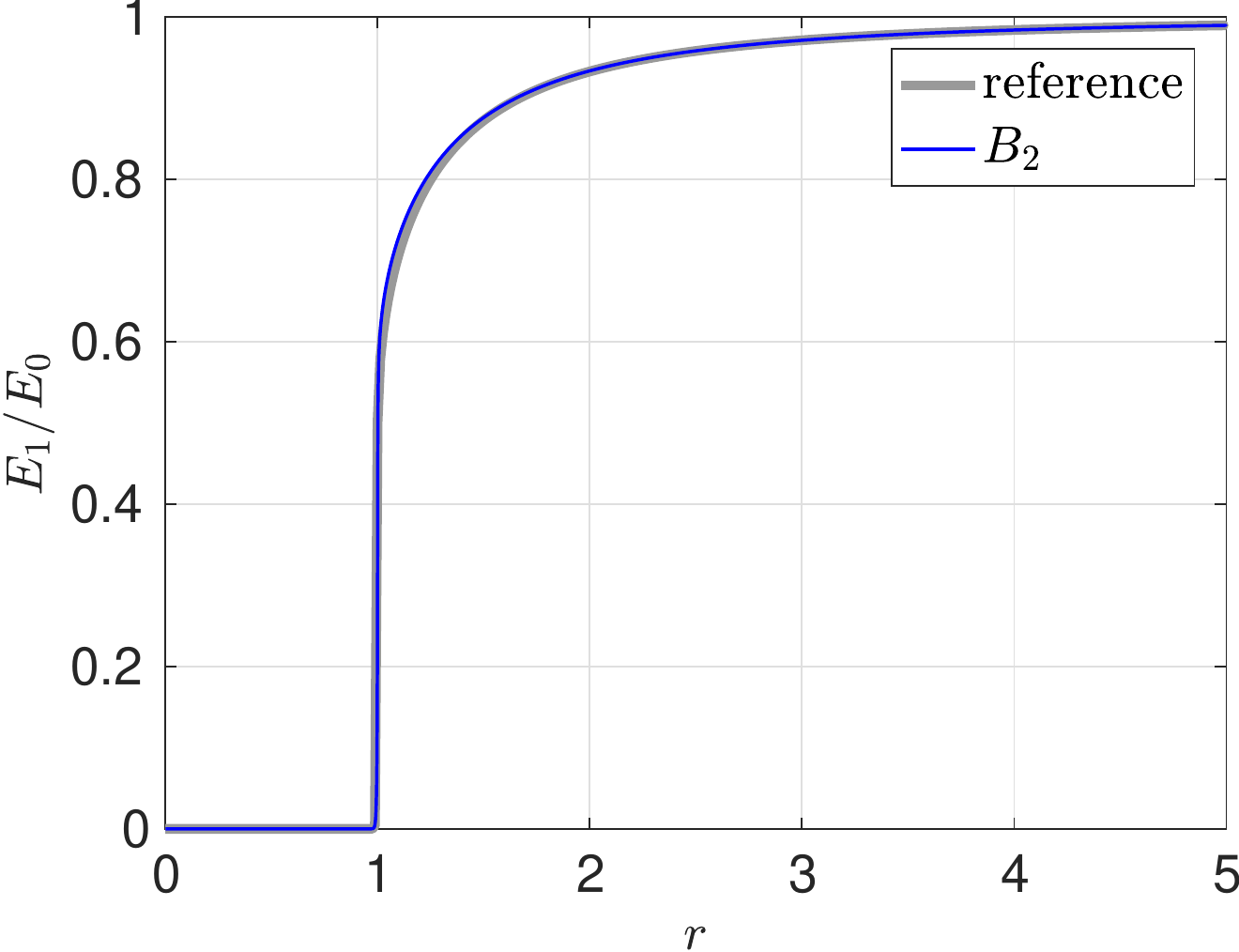}}
    \subfigure[Solution of $E_2/E_0$.]{
    \includegraphics[width=0.3\textwidth,height=0.16\textheight]{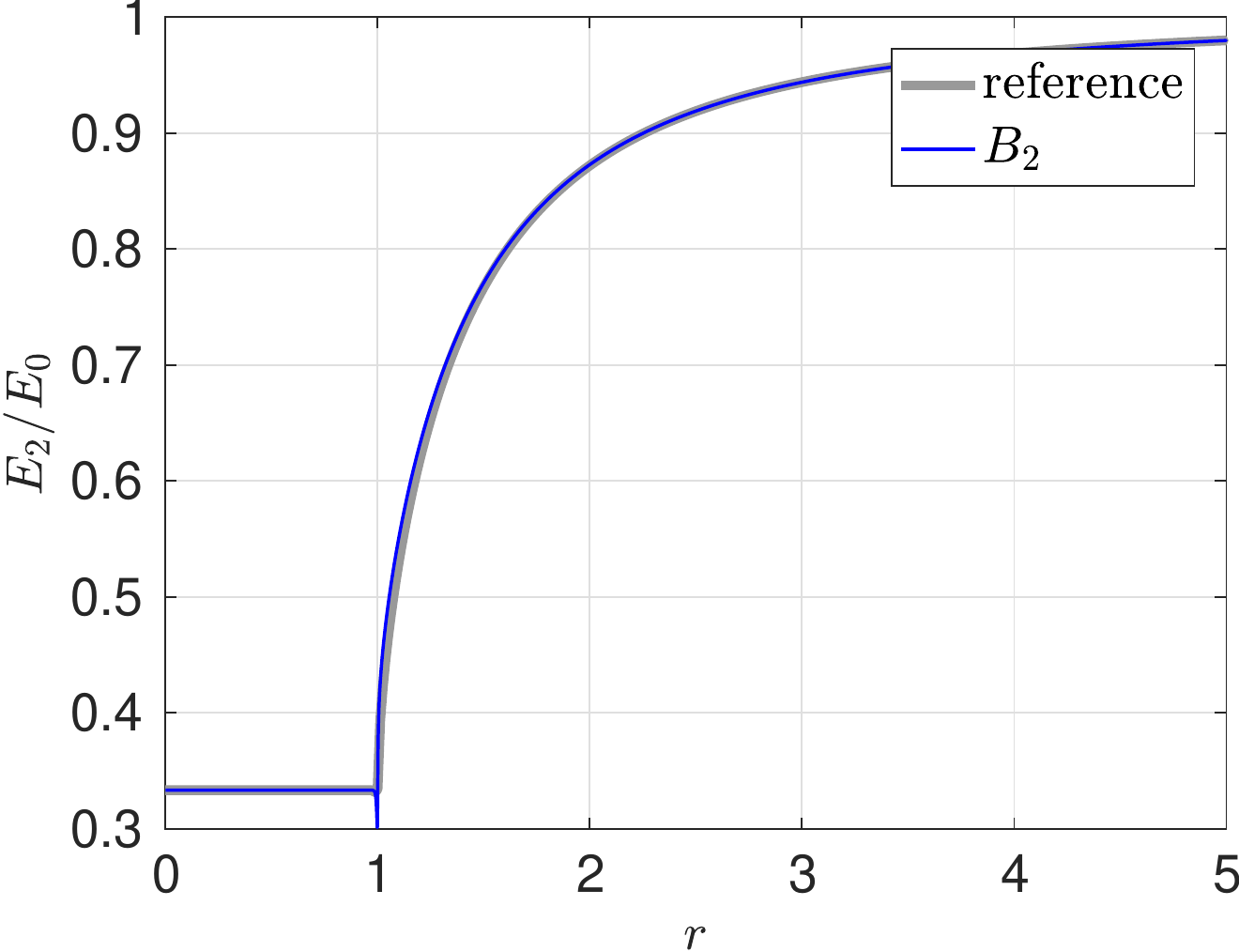}}
    \subfigure[Solution of $E_2/E_0-(E_1/E_0)^2$.]{
    \includegraphics[width=0.3\textwidth,height=0.16\textheight]{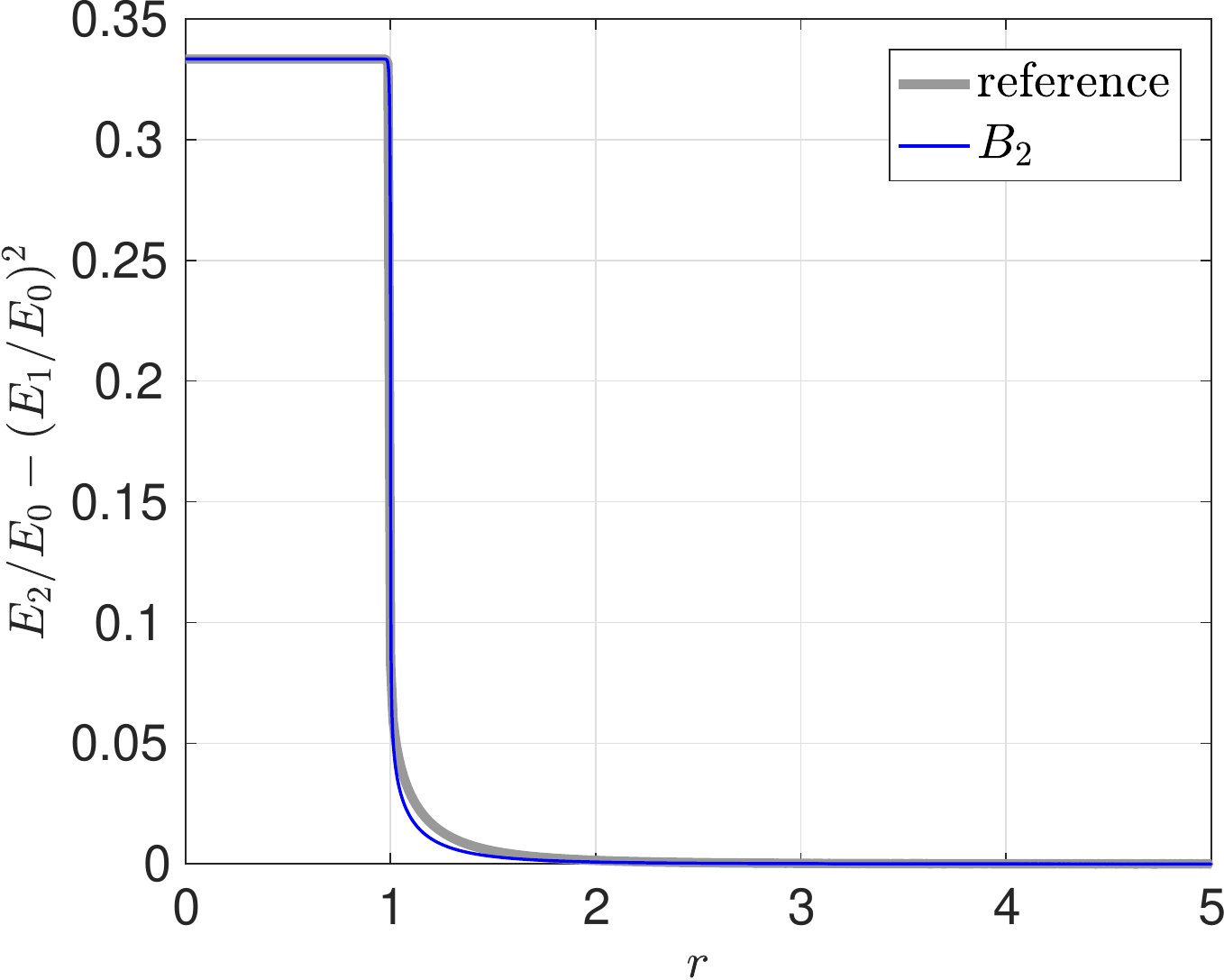}}
    \caption{Numerical results of the $B_2$ model for homogeneous sphere example}
    \label{fig:HomogeneousSphere}
\end{figure}

\begin{figure}[htbp]
    \centering
    \subfigure[Solution of $P_{10}$]{  
    \includegraphics[width=0.3\textwidth,height=0.16\textheight]{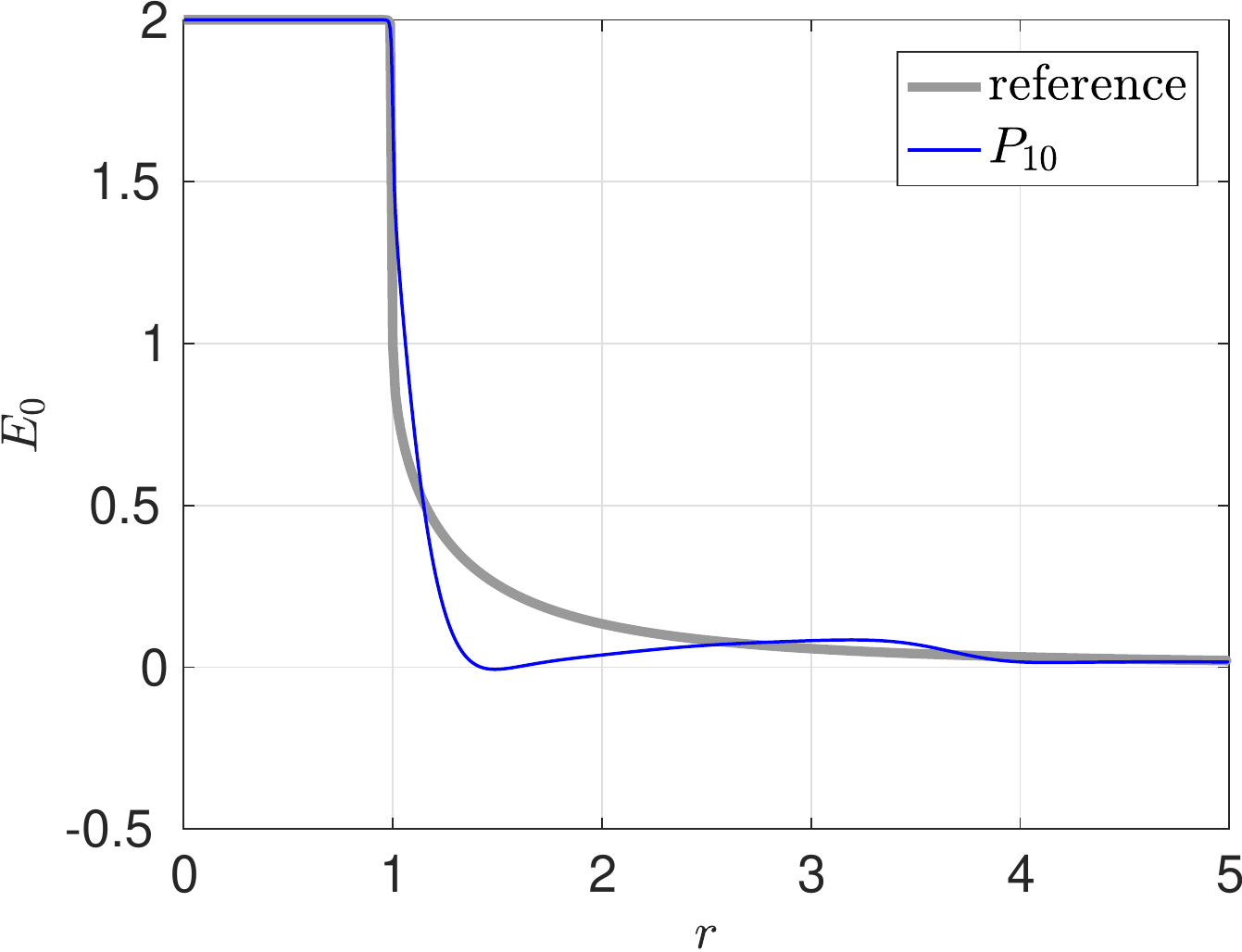}}
    \subfigure[Solution of $P_{30}$]{
    \includegraphics[width=0.3\textwidth,height=0.16\textheight]{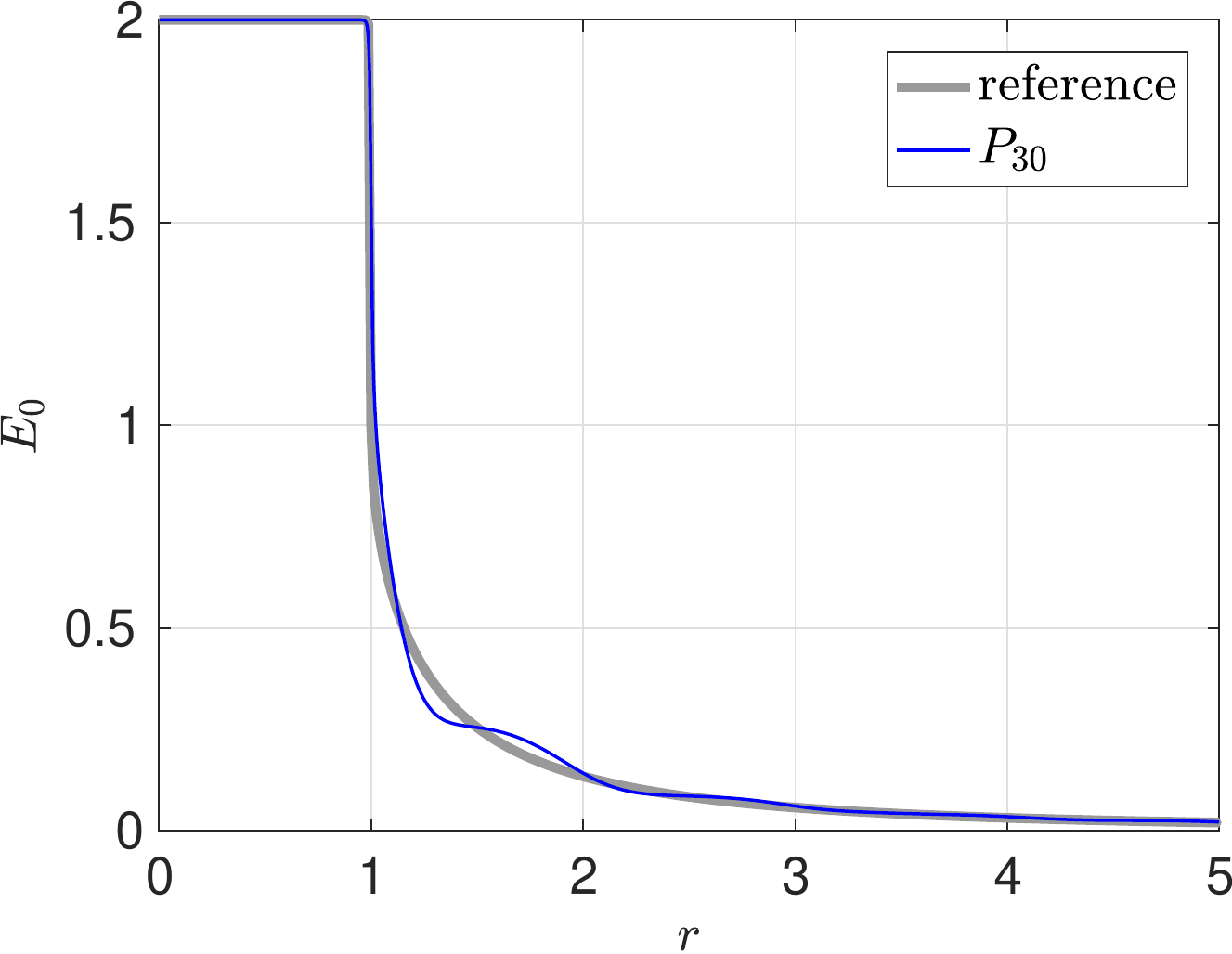}}
    \subfigure[Solution of $P_{50}$]{
    \includegraphics[width=0.3\textwidth,height=0.16\textheight]{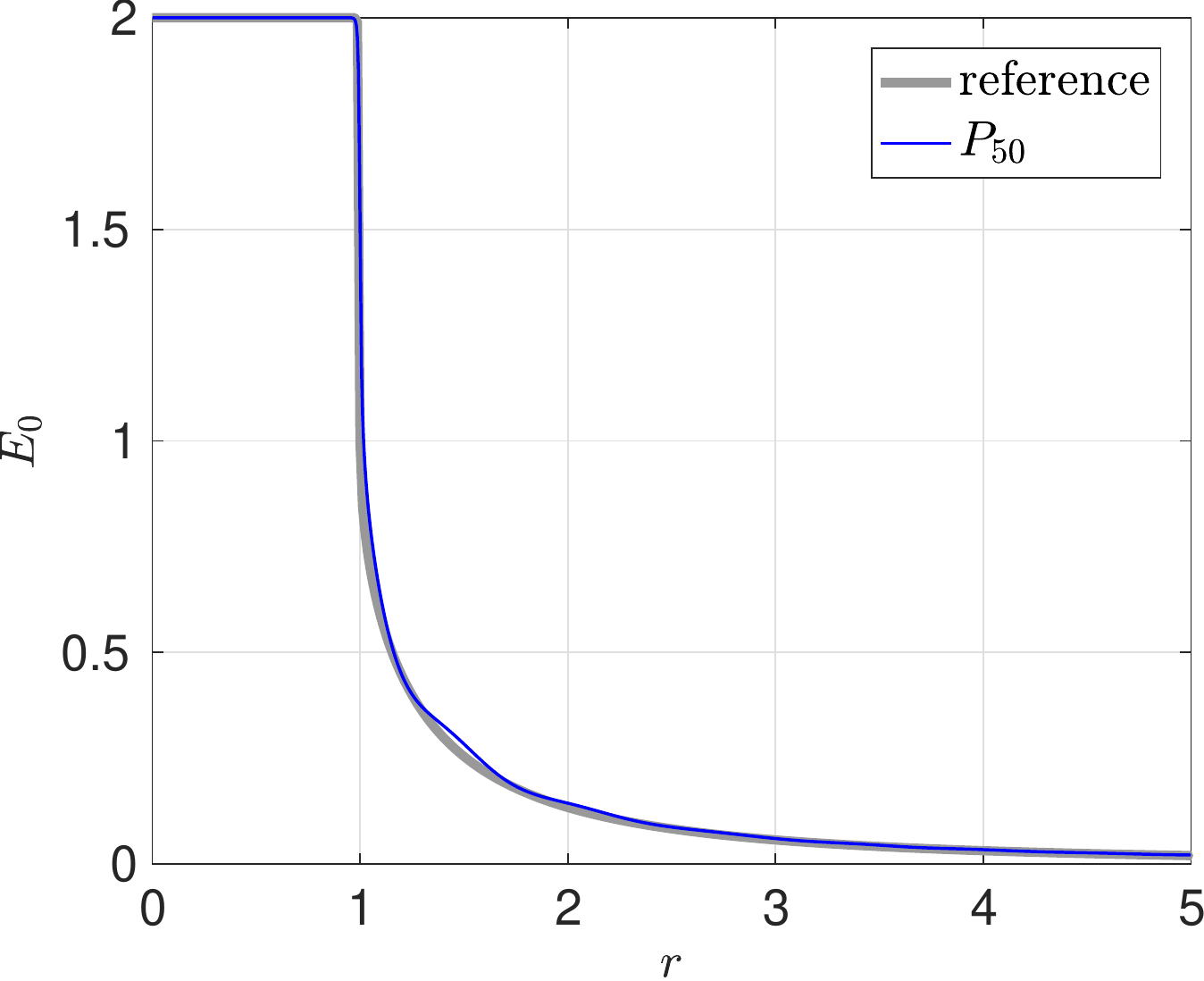}}
    \caption{Numerical results of the $P_N$ model for homogeneous sphere example}
    \label{fig:HomogeneousSpherePn}
\end{figure}

We use $N_{\text{cell}}=$ 
10000 cells to calculate this example and
the results of $E_0$, $E_1$, and $E_2$ of the $B_2$ model are shown in
\Cref{fig:HomogeneousSphere}. The results of the $P_2$, $P_6$, and
$P_{10}$ model are shown in \Cref{fig:HomogeneousSpherePn}. 
The values of $\frac{E_1}{E_0}$, $\frac{E_2}{E_0}$, and
$\frac{E_2}{E_0}-\left( \frac{E_1}{E_0} \right)^2$ in
\Cref{fig:HomogeneousSphere} gives some
indication as to the characteristics of the distribution function of
this problem.  First, we see that the distribution function of this
numerical solution lies in the realizable region for any $r$. Second,
when $r\leq R$, $\frac{E_1}{E_0}=0$ and $\frac{E_2}{E_0}=\frac{1}{3}$.
This indicates that the distribution function
$I(r,\mu)=\frac{E_0}{2}$, which is isotropic; when $r>R$, the results
show that the distribution function gets more anisotropic when $r$
gets larger. When $r$ is large, according to the result of
$\frac{E_1}{E_0}$ and $\frac{E_2}{E_0}-\frac{E_1^2}{E_0}$, we know
that the distribution function is close to the boundary of the
realizable region. The results of $B_2$ model in \Cref{fig:HomogeneousSphere} 
all agree well with the analytical
solution,
showing
that the $B_2$ model is valid to simulate both isotropic and
anisotropic distribution functions.


Furthermore, notice that when $r<R$, because $\sigma=250\gg 1$, the distribution
function of the analytical solution 
is close to a constant. However, when $r\geq R$, the
distribution function itself has a discontinuity with respect to $\mu$. 
Therefore, this problem not only has a discontinuity at the sphere with
respect to spatial variable $r$, but also has a discontinuity with
respect to velocity related variable $\mu$. Actually, it is this discontinuity 
which leads to the anisotropy of the distribution function.
This phenomenon, which is 
often found in many practical applications, 
is a challenge for numerical simulations 
\cite{abdikamalov2012new,Smit1997Hyperbolicity}.
However, the results of the $B_2$ model show that the $B_2$ model can
handle the discontinuity well. On the other hand, according to the
results of the $P_N$ model in \Cref{fig:HomogeneousSpherePn}, the
$P_N$ model can not handle this discontinuity, and we need a relative
large order to obtain a satisfying result.

\begin{example}[Static scattering atomsphere]
  This example is similar to test cases considered in
  \cite{hummer1971radiative,abdikamalov2012new}.
Consider the pure scattering steady-state radiative transfer equation
\begin{equation}
  \mu \pd{I}{r} + \dfrac{1-\mu^2}{r}\pd{I}{\mu} =
  -\sigma_s I + 
  \dfrac{\sigma_s}{2} \int_{-1}^1 I(\mu)\dd\mu.
\end{equation}
where the computational region is $r\in[0, R]$, $\sigma_s = 
r^{-n}$ with $n > 1$. 
A vacuum boundary condition is prescribed at the right boundary. 
In addition, a point source is located at the origin, which emits $L$
units of energy per unit time.  This is the simplest spherical
analogue of the classical Milne problem in the plane-parallel
geometry.
\end{example}
The analytical result of this problem was studied in
\cite{hummer1971radiative}, where the researchers show the asymptotic
behavior of $E_0$ when $r$ is sufficiently small and $r$ is
sufficiently large. The tendency of $E_0$ can be formulated as 
\begin{equation}
  \label{eq:milneproblemtendency}
E_0 \propto
\left\{
\begin{aligned}
   &r^{-(n+1)}, \quad & r\ll 1;\\
   &r^{-2},\quad & 1\ll r\leq R.
\end{aligned}
\right.
\end{equation}

\begin{figure}[htbp]
    \centering
    \subfigure[solution of $E_0$ when $n=1.5,R=1$]{  
    \includegraphics[width=0.3\textwidth,height=0.16\textheight]{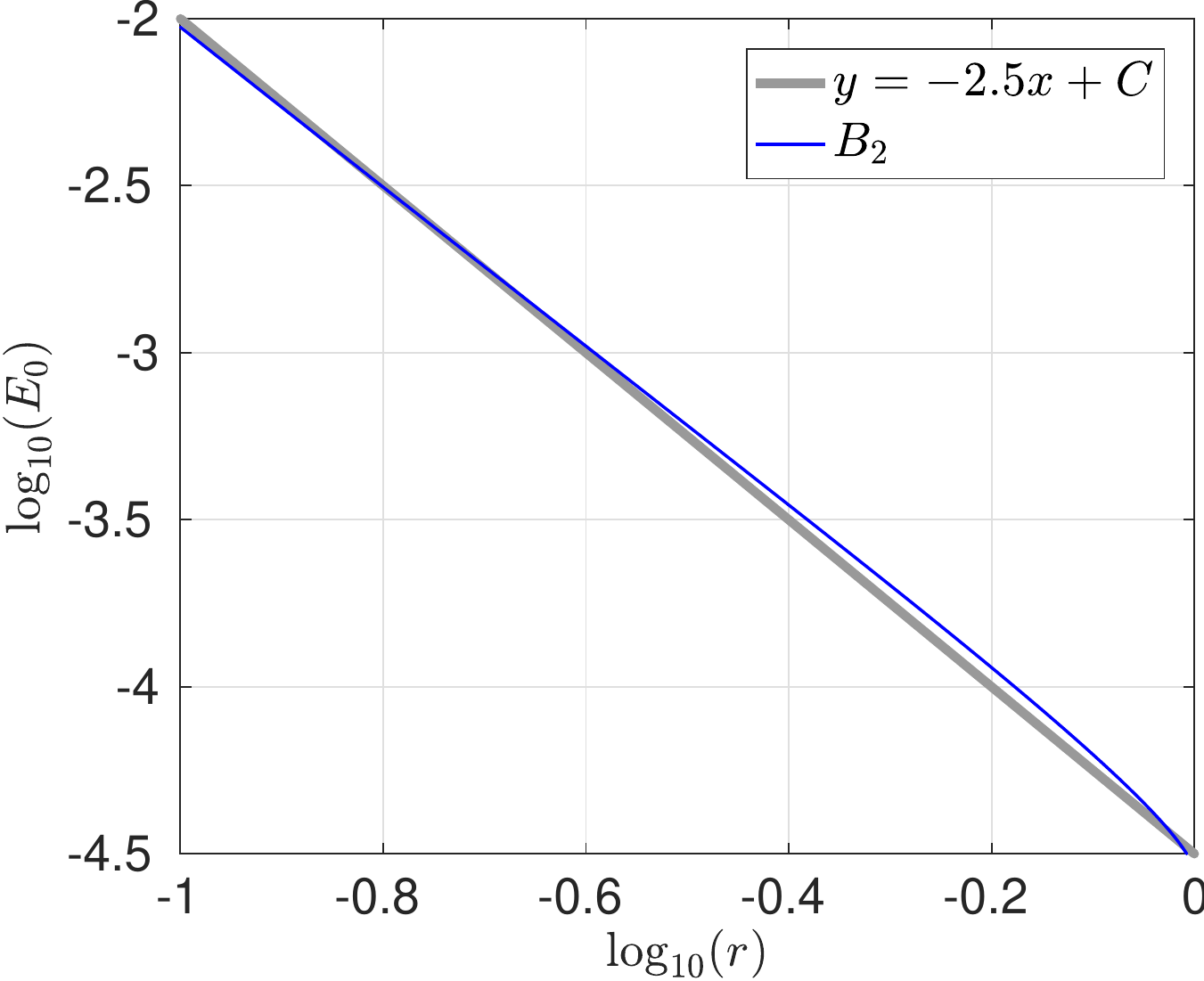}}
    \subfigure[solution of $E_0$ when $n=2,R=1$]{
    \includegraphics[width=0.3\textwidth,height=0.16\textheight]{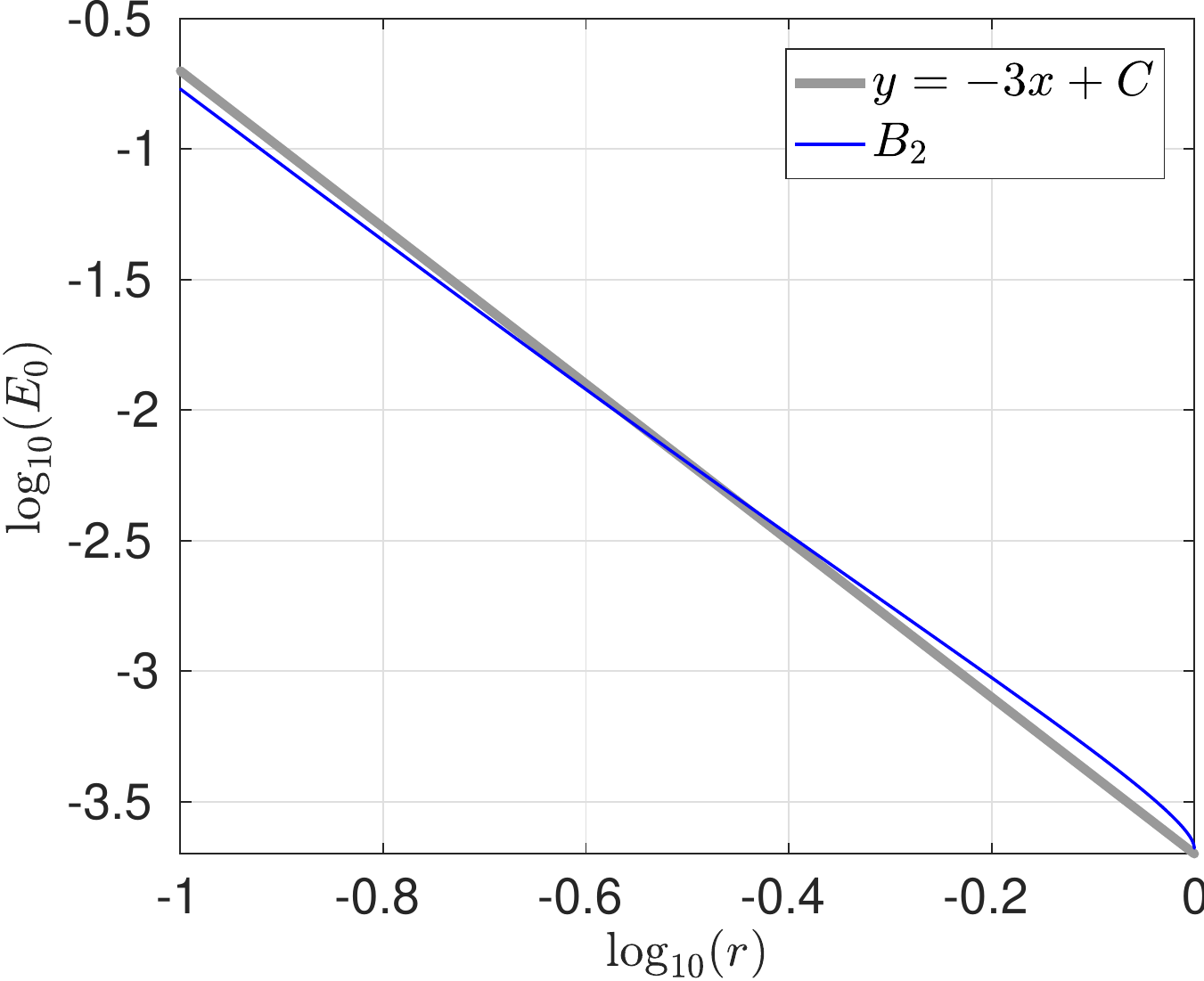}}
    \subfigure[solution of $E_0$ when $n=3, R=1$]{
    \includegraphics[width=0.3\textwidth,height=0.16\textheight]{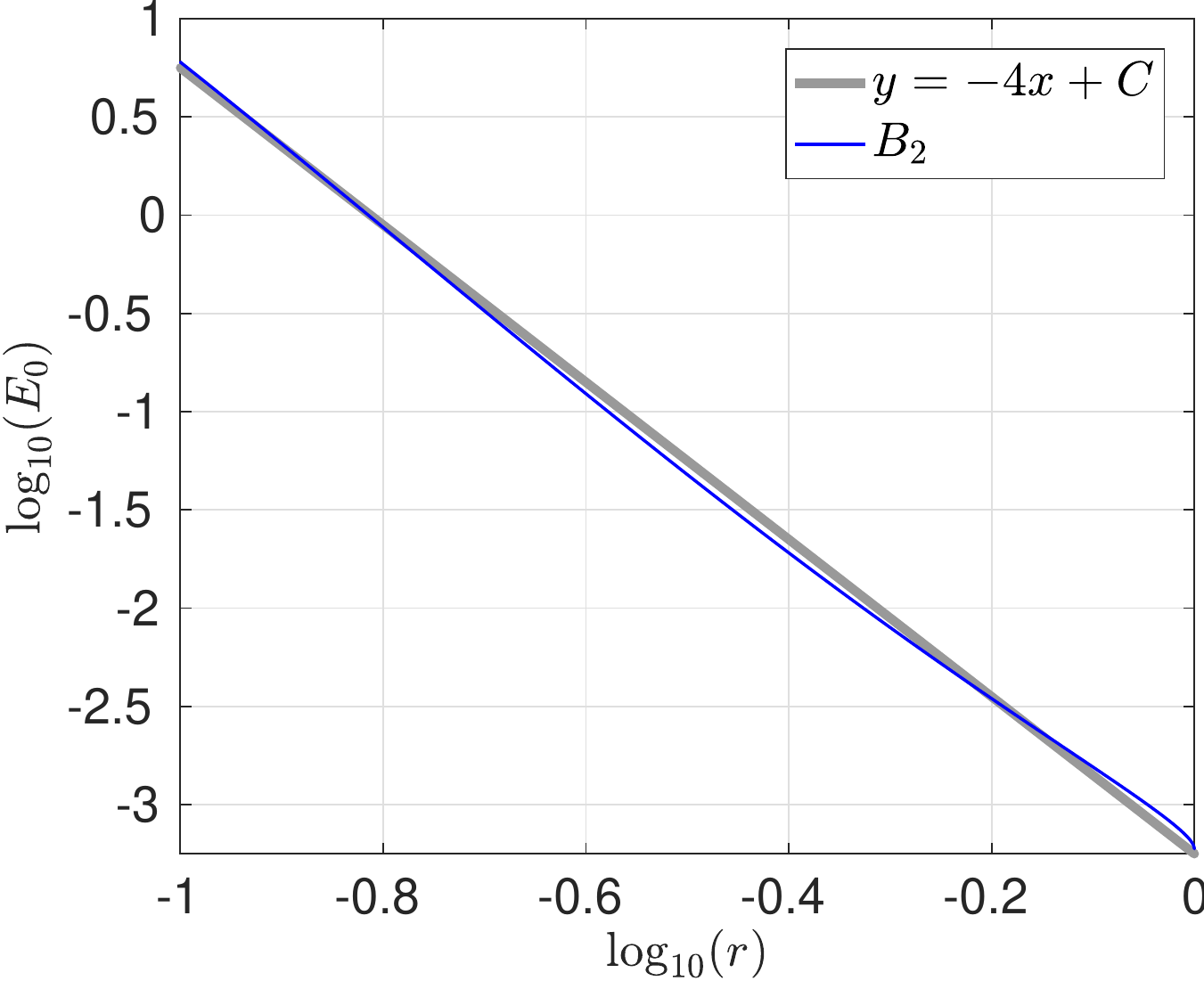}}
    \hfill
    \subfigure[solution of $E_0$ when $n=1.5, R=100$]{
    \includegraphics[width=0.3\textwidth,height=0.16\textheight]{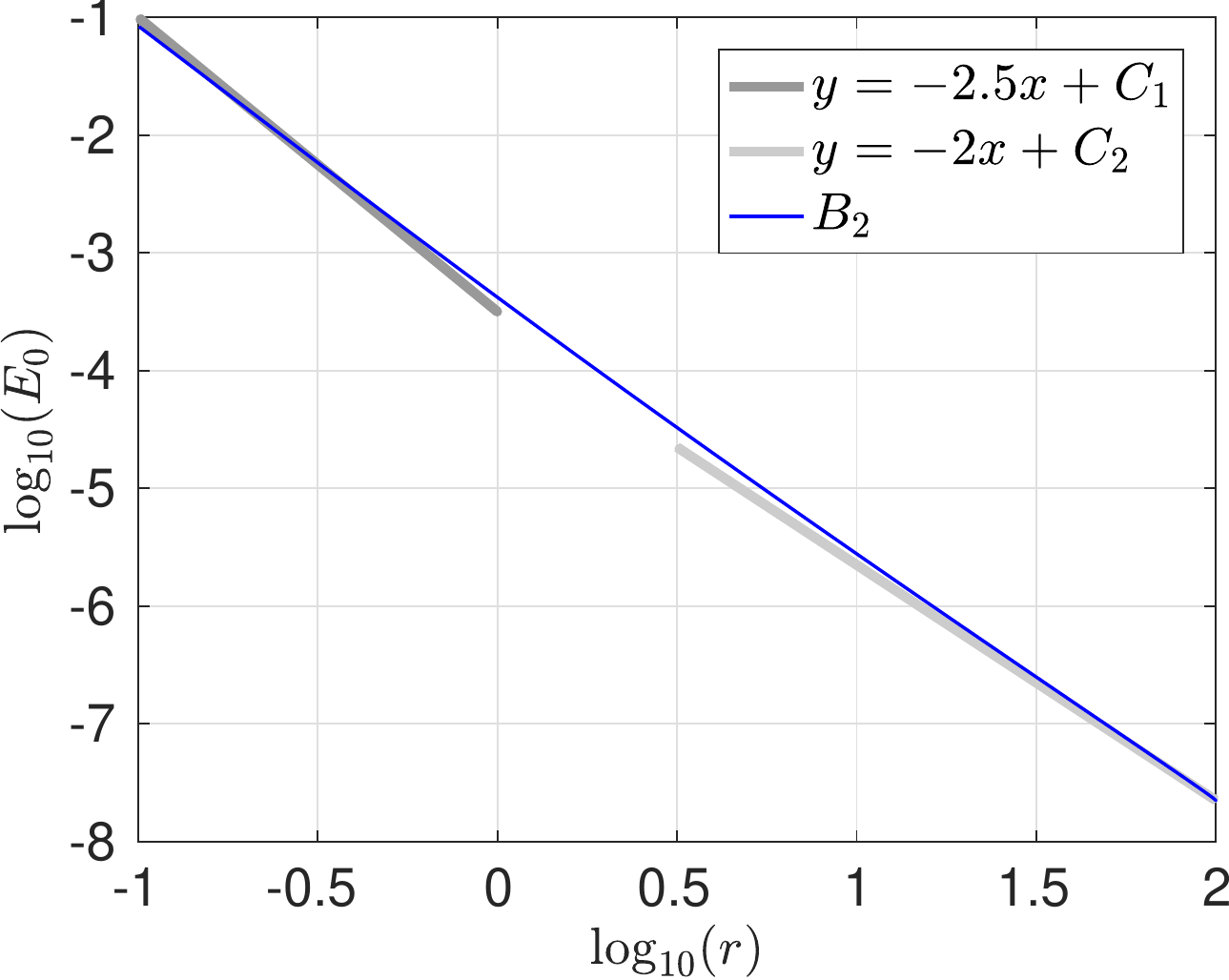}}
    \subfigure[solution of $E_0$ when $n=2, R=100$]{
    \includegraphics[width=0.3\textwidth,height=0.16\textheight]{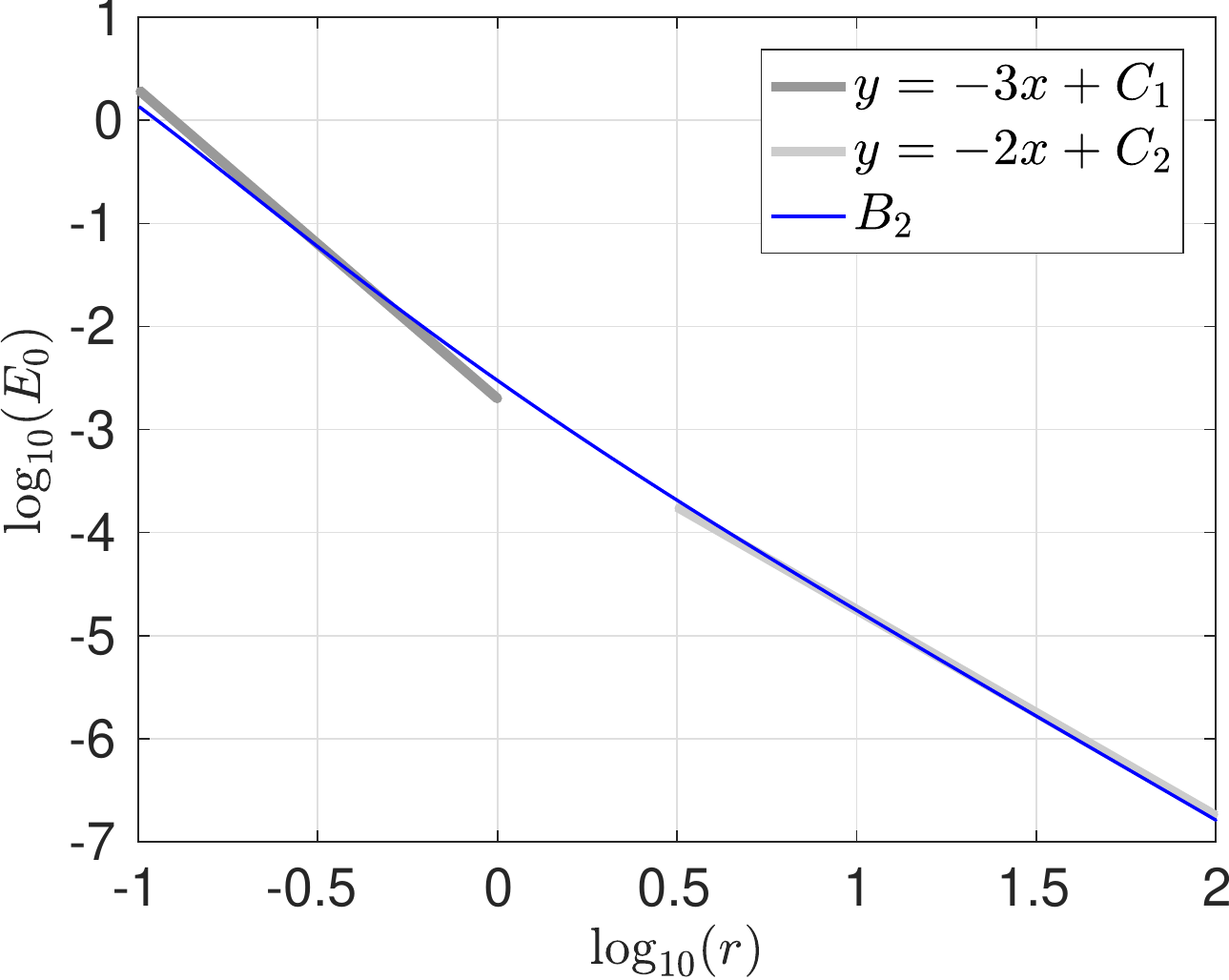}}
    \subfigure[solution of $E_0$ when $n=3, R=100$]{
    \includegraphics[width=0.3\textwidth,height=0.16\textheight]{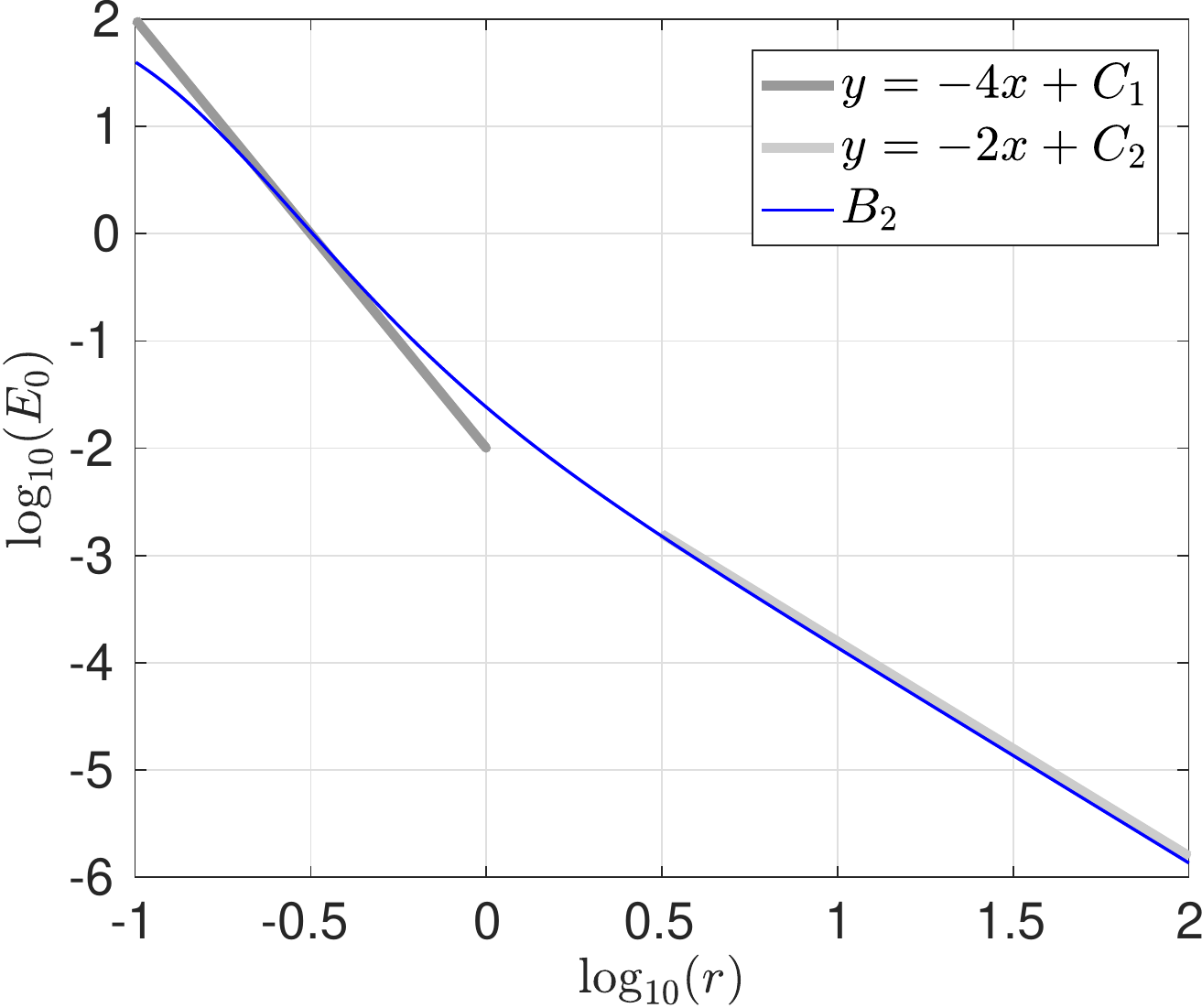}}
    \caption{Numerical results for static scattering atomsphere.}
    \label{fig:Milneproblem}
\end{figure}

We perform a set of simulations with $L =$ 1, $R =$ 1 and 100, and $n
= \frac{3}{2}$ , 2, and 3. When $R=1$, $N_{\text{cell}}=10000$ cells
is used, and when $R=100$, we use $N_{\text{celll}}=100000$ cells.  In
\Cref{fig:Milneproblem} we show the results of $E_0$ and the
comparison with lines to determine the slopes.  The numerical result
for $E_0$ shown in \Cref{fig:Milneproblem} is with the logarithm
scale.  Notice that
$C$, $C_1$, and $C_2$ can be different in different subfigures.
According to the results for different choices of $n$ in
\Cref{fig:Milneproblem}, we know that the results of the $B_2$ model
is consistent with the asymptotic analysis in
\eqref{eq:milneproblemtendency}.

Notice that in this example, $E_0$ tends to infinity when the radius $r$
tends to 0. This means that in the geometry source term
$\dfrac{1-\mu^2}{r}\pd{I}{\mu}$ is extremely stiff,
which is a challenge in numerical simulations.
However, the $B_2$ model yields correct result
with a CFL number equals to 0.95. Therefore,  
the $B_2$ model and the corresponding numerical
scheme we adopt are valid and robust to simulate the RTE.

\section{Conclusion}\label{sec:conclude}
We extended the $B_2$ model for the radiative transfer equation in slab
geometry, which was studied in \cite{alldredge2016approximating}, to
the spherically symmetric case. The global hyperbolicity and
realizability of the $B_2$ model can be verified, and its
characteristic structure was analyzed in detail. One of the main
advantages of the $B_2$ model is that its ansatz could capture highly
anisotropic distribution functions, while at the same time it has
explicit closing relationship. In our test cases where there are
strongly peaked or discontinuous distribution functions the $B_2$
model shows better performance than the $P_n$ model, displaying its
potential to approximate the RTE accurately and inexpensively. The
next step is to extend the model to higher order.

\section*{Acknowledgements}
The work of R.L. and L.Z. is partially supported by Science Challenge
Project, No. TZ2016002 and the National Natural Science Foundation of
China (Grant No.  91630310 and 11421110001, 11421101). The work of
W.L. is supported by NSFC Grant No. U1530401. The computational
resources are supported by the high performance computing platform of
Peking University, China.

\bibliographystyle{plain}
\bibliography{../references}

\end{document}